\documentclass[a4paper,11pt]{article}
\pdfoutput=1 	% if your are submitting a pdflatex (i.e. if you have
\usepackage{jheppub} 	% for details on the use of the package, please
\usepackage{verbatim}
\usepackage{graphicx}	% Include figure files
\usepackage{dcolumn}	% Align table columns on decimal point
\usepackage{bm}		    % bold math
\usepackage{hyperref}
\usepackage{amsmath}
\usepackage{verbatim}
\usepackage{slashed}
\usepackage{hyperref}
\usepackage{mathtools}
\usepackage{tabularx}
\usepackage{xfrac}
\usepackage{multirow}
\usepackage{comment}
\usepackage{ytableau}
\ytableausetup{boxsize=0.6em}
\usepackage{amsmath}

\preprint{IBS-CTPU-26-29}

\title{\boldmath  Constructing $SU(N)$ strongly coupled dark sectors by extending the Standard Model: mixed representations, flavor structure, global symmetries and various portals
}

\author{Yi Chung}

\affiliation{
Particle Theory and Cosmology Group, Center for Theoretical Physics of the Universe, \\Institute for Basic Science (IBS), Daejeon, 34126, Korea
}

\emailAdd{yichung886@ibs.re.kr}

\abstract{Motivated by the potential existence of a dark QCD sector, we construct a $SU(N)$ strongly coupled dark sector by extending the Standard Model (SM) gauge group to an enlarged $SU(N+5)\times SU(N+1)\times U(1)$, under which the dark quark candidates naturally arises from the same multiplets as the SM fermions. The dark QCD sector from the extension inherits the rich structure of the SM, featuring three distinct properties: (1) Mixed representations: There are dark quarks of three different representations, including fundamental ($\chi_f$), antisymmetric ($\chi_a$), and symmetric ($\chi_s$), which can be traced back to the chiral structure of the SM. (2) Flavor: Each representation features three generations, with their flavor structures and Yukawa interactions directly related to those of the SM. (3) Global symmetries: Alongside the conventional dark baryon number $D$, we identify an exotic dark baryon number $\widetilde{D}$ from a conserved $(B-L)-\widetilde{D}$ symmetry. These features provide novel perspectives on the structure of dark QCD sectors. Furthermore, the extension of the scalar and gauge sectors introduces various portals between the dark QCD sector and the SM sector. We systematically study these portals, including quark portal, lepton portal, Higgs and $Z$ portal, heavy $Z'$ portal, kinetic mixing portal, dipole portal, and dark pion/axion portals. With a concrete UV model, their origin become robust, providing additional guidance for experimental searches of the dark sector.}

\begin{document}
\maketitle
\flushbottom

%\newpage

\section{Motivation}

%~\\ - Dark Matter ~\\

The Standard Model (SM) of particle physics successfully describes all known elementary particles and their interactions. However, several observations remain unexplained, particularly those arising from cosmology and astrophysics, including the existence of dark matter, dark energy, and the origin of the matter-antimatter asymmetry. Among these open questions, the nature of dark matter is one of the earliest recognized mysteries , yet still remains one of the most challenging problems in particle physics. Among the various dark matter candidates, the existence of a strongly coupled dark sector \cite{Strassler:2006im,Kribs:2016cew,Asadi:2026mip} offers several attractive features but remains relatively unexplored, motivating further study.

%~\\ - Dark QCD sector ~\\

A strongly coupled dark sector typically features a non-Abelian gauge group that becomes strongly coupled at an infrared scale. Depending on the matter content charged under the gauge group, the resulting low-energy spectrum can be quite diverse, featuring different dark matter candidates, such as dark baryons \cite{Nussinov:1985xr,Chivukula:1989qb,Barr:1990ca,Francis:2018xjd}, dark pions \cite{Bhattacharya:2013kma,Hochberg:2014dra,Hochberg:2014kqa,Alfano:2025non}, or dark glueballs \cite{Okun:1980mu,Faraggi:2000pv,Juknevich:2009gg}. Their masses are dynamically generated by the dark gauge dynamics, and their stability is guaranteed by accidental symmetries or highly suppressed decay rates. It is worth noting that the most abundant visible particle is the proton, whose mass is generated by quantum chromodynamics (QCD) and whose stability is ensured by accidental baryon number conservation. The dark sector can therefore exhibit a similar structure.

%~\\ - Dark Matter - Baryon Coincidence ~\\

The relic abundance of a strongly coupled dark sector, depending on the details of the complete model and its dark matter candidate, can originate from diverse cosmological processes but should always reproduce the correct dark matter energy density, which has been measured precisely \cite{Planck:2018vyg,AtacamaCosmologyTelescope:2025blo,SPT-3G:2025bzu}. The production mechanism can either be a conventional thermal process, such as freeze-out and freeze-in, or an asymmetric generation mechanism analogous to the baryogenesis of QCD baryons. The latter one, giving rise to the so-called Asymmetric Dark Matter (ADM), is especially attractive as it predicts comparable number densities, $n_B\approx n_D$, between QCD baryons and dark matter \cite{Kaplan:2009ag,Davoudiasl:2012uw,Petraki:2013wwa,Zurek:2013wia}, providing part of a solution to the intriguing $\mathcal{O}(1)$ ratio between the baryon and dark matter abundances \cite{Turner:1987pp}. Moreover, a complete solution, which also explains comparable masses, $m_B\approx m_D$, usually requires dark matter to be a dark baryon within a strongly coupled dark sector \cite{Hodges:1993yb,Berezhiani:1995am,Foot:2003jt,An:2009vq,Farina:2015uea,GarciaGarcia:2015pnn,Lonsdale:2018xwd,Lonsdale:2014wwa,Lonsdale:2014yua,Ibe:2019ena,Rosa:2022sym,Bodas:2024idn,Cox:2025wxk,Bai:2013xga,Newstead:2014jva,Ritter:2022opo,Ritter:2024sqv,Murgui:2021eqf,Chung:2024ezq,Chung:2024nnj}, providing a further motivation for this class of dark sector models.

%~\\ - Dark Grand Unification as UV of ADM ~\\

However, to realize the mechanism of asymmetric dark matter, or more generally to construct a viable dark sector model, heavy states or higher-dimensional operators that connect the dark sector to the visible sector are required, sometimes featuring certain symmetries. However, in most studies, these heavy states and interaction terms are introduced by hand without clear guidance from a UV completion. The issue can be further extended to the structure of the dark sector itself. Therefore, in this study, we construct a strongly coupled dark sector by extending the unified SM gauge group $SU(5)\times U(1)_X$ to $SU(N+5)\times SU(N+1)' \times U(1)''$, following previous attempts \cite{Barr:2011cz,Barr:2013tea,Barr:2015lya,Chung:2024nnj} but within a more general framework, yielding an $SU(N)$ dark gauge group. We then focus on the $SU(N)$ dark sectors that can become strongly coupled in the infrared.

%\newpage

%~\\ - More about dark sector features ~\\

The extension of the matter content associated with the extended gauge group determines all the ingredients of the dark sector. From the extension of the fermion multiplets, new SM singlets naturally arise, which can play the role of dark quarks in the model. Since the SM fermions are arranged in different representations under the $SU(5)\times U(1)_X$, the dark sector also features mixed fermion representations, introducing a rich dark QCD sector that has seldom been considered before. Besides, the dark quarks also feature three generations and a specific flavor structure, just like the SM fermions.

On the other hand, the extension of the scalar and gauge sectors introduces heavy particles that serve as mediators, resulting in various portal interactions, which are categorized systematically in this study. Since some of the Yukawa couplings arise from the same Yukawa interactions that give rise to the SM fermion masses, we can even roughly predict their magnitudes and flavor structures, providing additional guidance for the phenomenology study. Furthermore, the symmetry structure behind these interaction terms also becomes clear, as it is no longer introduced by hand but emerges from the gauge structure in the far UV, allowing us to identify new global symmetries, including $(B-L)-\widetilde{D}$ with exotic dark baryon number $\widetilde{D}$, which we identify for the first time in this study.

%~\\ - Other applications ~\\

Although we focus on the construction of dark sectors in this work, we expect such an extension to be generally useful for building strongly coupled hidden sectors beyond the SM for other purposes. In this work, we focus on the SM singlets arising from the extension that can serve as dark matter candidates. However, there are also exotic fermions carrying both $SU(N)$ color charges and SM charges, which can, with electroweak-charged fermions, realize fundamental composite Higgs models  \cite{Galloway:2010bp,Barnard:2013zea,Ferretti:2013kya,Ferretti:2014qta,Cacciapaglia:2014uja,Cacciapaglia:2020kgq,Chung:2021ekz,Chung:2021xhd} or, with color-charged fermions, realize composite axion models \cite{Kim:1984pt,Choi:1985cb,Kaplan:1985dv,Randall:1992ut,Dobrescu:1996jp,Redi:2016esr,Lillard:2017cwx,Lillard:2018fdt,Contino:2021ayn,Gherghetta:2025fip,Gherghetta:2025kff,Azatov:2025mep}. 

%~\\ - Organization of the paper ~\\

This paper is organized as follows. We start with the construction of the model from the bottom up in section \ref{sec:Idea}, explaining how the SM is extended. Given the complete gauge group and fermion sector, we then proceed from the top down to include the scalar sector and set up the required symmetry breaking in section \ref{sec:Model}. The relevant Yukawa interactions and fermion mass terms are presented in detail, along with a discussion of the accidental $U(1)$ symmetries. With all the ingredients presented, we analyze the resulting $SU(N)$ strongly coupled dark sector in section \ref{sec:DS}, with subsections for different values of $N$. Next, in section \ref{sec:Portal}, we systematically present all the possible portal terms arising from the heavy mediators in this model, including a brief discussion of their phenomenological implications and direct searches. Section~\ref{sec:Conclusion} concludes the paper.

%\newpage

\section{The Idea}\label{sec:Idea}

To construct a dark sector by extending the SM, we would like to first arrange the SM gauge group and fermion content under a unified framework. This step is not necessary but is helpful for keeping track of the symmetry structure, especially global symmetries such as $B-L$. It was also pointed out that an extension based on the Georgi-Glashow $SU(5)$ Grand Unified Theory (GUT) \cite{Georgi:1974sy} can naturally lead to additional SM singlet fermions that can serve as dark quarks \cite{Barr:2011cz,Barr:2013tea,Barr:2015lya,Chung:2024nnj}.\footnote{For other unified theories, such as the Pati-Salam model \cite{Pati:1974yy}, the extension typically involves only SM-charged fermions and therefore requires the introduction of additional multiplets as dark quarks \cite{Murgui:2021eqf}.} In this study, we further extend the previous idea to construct $SU(N)$ strongly coupled dark sectors.

\subsection{The fermion sector from the extension of the SM}

Starting by unifying the SM fermion content under the $SU(5)$ GUT framework, the fermion content of the SM can be embedded into an antisymmetric rank-two representation and an anti-fundamental representation, as
\begin{align}
\mathbf{10}_F=
\begin{pmatrix}
u^c     &  q  \\
-q^T     &  e^c  \\
\end{pmatrix}
~\text{ and }~
\overline{\mathbf{5}}_F=
\begin{pmatrix}
d^c \\ \ell 
\end{pmatrix}~\text{ for each generation,}
\label{SU(5)F}
\end{align}
where we use the Weyl spinor notation following the conventional GUT studies. Next, we extend the $SU(5)$ GUT gauge group to $SU(N+5)$ in order to include an $SU(N)$ dark color group. The fermion content is automatically determined by this extension, and is given by
\begin{align}
\mathbf{A}^{[\alpha\beta]}=
\begin{pmatrix}
\chi_a      &   \psi_c   &  \psi_w \\
-\psi_c^T   &   u^c      &  q      \\
-\psi_w^T   &   -q^T     &  e^c    \\
\end{pmatrix}
~\text{ and }~
\overline{\mathbf{F}}_\alpha=
\begin{pmatrix}
\chi_f^c \\ d^c \\ \ell 
\end{pmatrix}~\text{ for each generation,}
\label{SU(N+5)F}
\end{align}
where $\alpha,\beta=1,2...N+5$ are $SU(N+5)$ indices. Four new types of dark fermions are introduced, labeled by $\chi$ and $\psi$ with different subscripts. They transform under the gauge group $SU(N) \times SU(3)_C \times SU(2)_W \times U(1)_Y$ as
\begin{align}\label{chi}
\chi_a= (A,1,1)_0~,~\psi_c=(N,3,1)_{-\frac{1}{3}}~,~\psi_w =(N,1,2)_{\frac{1}{2}}~,~ \chi_f^c= (\bar{N},1,1)_0~.
\end{align}
The subscripts reflect their quantum numbers, with $a$ and $f$ denoting the antisymmetric and fundamental representations under $SU(N)$, respectively, and with $c$ and $w$ indicating that they carry charges under the SM $SU(3)_C$ and $SU(2)_W$, respectively. From the new fermion content, we can identify two SM-singlet fermions, $\chi_a$ and $\chi_f^c$, that can serve as dark quark candidates, showing one of the advantages of extending the $SU(5)$ GUT to include the dark color group as discussed above.

However, this cannot be the complete fermion content, since $SU(N+5)$ is anomalous. To cancel the anomaly, $N$ additional fermions in the anti-fundamental representation, $\overline{\mathbf{F}}_\alpha$, are required, which are written as
\begin{align}
N \times \overline{\mathbf{F}}_\alpha=
\begin{pmatrix}
\psi_f^c \\ \psi_c^c \\ \psi_w^c 
\end{pmatrix}~\text{ for each generation.}
\label{NFbar}
\end{align}
In this new multiplet, there are three new types of fermions, all labeled by $\psi$ with different subscripts, following the labeling convention introduced above. They transform under the group $SU(N)' \times SU(N) \times SU(3)_C \times SU(2)_W \times U(1)_Y$ as
\begin{align}\label{psi}
\psi_f^c= (\bar{N},\bar{N},1,1)_0~,\quad \psi_c^c=(\bar{N},1,\bar{3},1)_{\frac{1}{3}}~,\quad \psi_w^c =(\bar{N},1,1,\bar{2})_{-\frac{1}{2}}~,
\end{align}
where $SU(N)'$ is the global symmetry acting on the $N$ new anti-fundamental fermions. We choose them to transform as $\bar{N}$ under $SU(N)'$ for later convenience. There are again new SM-singlet fermions, $\psi_f^c$, that could serve as dark quarks.

However, the extension not only introduces SM-singlet fermions but also SM-charged fermions, which, if massless, would immediately render the model unrealistic. There are dark fermions $\psi_c$ and $\psi_w$ from $\mathbf{A}$, as well as $\psi_c^c$ and $\psi_w^c$ from $N \times \overline{\mathbf{F}}_\alpha$. These exotic SM-charged fermions should acquire masses large enough to evade current constraints. From their quantum numbers, however, it is impossible to write down gauge-invariant mass terms for these fermions. So how can these exotic fermions acquire sufficiently large masses?

It was already shown in \cite{Chung:2024nnj} that, if we gauge the $SU(N)'$ global symmetry and instead identify the diagonal subgroup of $SU(N)\times SU(N)'$ as the dark color gauge group $SU(N)_D$, mass terms between $\psi^c$ and $\chi$ can then be written down, as their quantum numbers under the SM gauge group match perfectly. Under this new setup, the fermions in eqs. \eqref{chi} and \eqref{psi} transform under the gauge group $SU(N)_D \times SU(3)_C \times SU(2)_W \times U(1)_Y$ as
\begin{align}\label{newfermion} 
&\chi_a= (A,1,1)_0~,~\psi_c=(N,3,1)_{-\frac{1}{3}}~,~\psi_w =(N,1,2)_{\frac{1}{2}}~,~ \chi^c_f= (\bar{N},1,1)_0~,\nonumber\\
&\psi_f^c=\chi_s^c\oplus\chi_a^c= (\bar{S},1,1)_0\oplus (\bar{A},1,1)_0~,~ \psi_c^c=(\bar{N},\bar{3},1)_{\frac{1}{3}}~,~ \psi_w^c =(\bar{N},1,\bar{2})_{-\frac{1}{2}}~,
\end{align}
where the fermions $\psi_s^c$ and $\psi_a^c$ come from the decomposition of $\psi_f^c$. We then find that there are three pairs of Weyl fermions that can form Dirac fermions, $(\chi_a,\psi_a^c)$, $(\psi_c,\psi_c^c)$, and $(\psi_w,\psi_w^c)$, thereby avoiding the issues associated with massless SM-charged Weyl fermions.

Next, we further organize all the anti-fundamental fermions into a single multiplet as
\begin{align}
\overline{\mathbf{F}}_{\alpha,\alpha'}=
\begin{pmatrix}
\psi_f^c=\chi_s^c\oplus\chi_a^c~ & \chi_f^c \\ \psi_c^c & d^c \\ \psi_w^c &  \ell
\end{pmatrix}~\text{ for each generation},
\label{N+1Fbar}~
\end{align}
and gauge the enlarged global symmetry $SU(N+1)'$ with index $\alpha'=1,2,...N+1$. Similarly, gauging the additional $SU(N+1)'$ group comes with the cost of additional gauge anomalies. Again, additional fermions are required to cancel the gauge anomalies. Following the study in \cite{Barr:2015rxa}, the minimal fermion content required to cancel the gauge anomaly is to introduce symmetric rank-two representations under the enlarged $SU(N+1)'$, which are given by
\begin{align}\label{fermionS}
\mathbf{S'}^{\{\alpha'\beta'\}}=
\begin{pmatrix}
\chi_s     &  \chi_f    \\
\chi_f^T   &  \psi_0    \\
\end{pmatrix}
~\text{ for each generation.}
\end{align}
The new fermions have the quantum number given by
\begin{align}\label{newfermion'} 
\chi_s=(S',1,1)_0~,\quad \chi_f =(N',1,1)_0~,\quad \psi_0=(1,1,1)_0
\end{align}
under $SU(N+1)' \times SU(3)_C \times SU(2)_W \times U(1)_Y$. A new set of SM-singlet fermions arises, which can again serve as dark quark candidates, except for $\psi_0$, which is chargeless under the dark color group. This fermion field, however, can be identified as the right-handed (RH) neutrino $\nu^c$, as we will see after identifying the $B-L$ symmetry. Therefore, this set of new fermion fields is not merely an ad hoc set of fermions added to cancel the anomaly but is in fact the extended part of the RH neutrinos, which is required if we want to realize the Type-I seesaw mechanism for neutrino masses or to gauge the $B-L$ symmetry.

\paragraph{The complete extended fermion sector}~\\

After the step-by-step construction of the extended fermion sector, we collect all the fermion content here. For each generation, there are fermions in three different representations, given by
\begin{align}
\mathbf{A}^{[\alpha\beta]}=
\begin{pmatrix}
\chi_a      &   \psi_c   &  \psi_w \\
   &   u^c      &  q      \\
   &        &  e^c    \\
\end{pmatrix}~,\quad
\overline{\mathbf{F}}_{\alpha,\alpha'}=
\begin{pmatrix}
\chi_s^c\oplus\chi_a^c & \chi_f^c \\ \psi_c^c & d^c \\ \psi_w^c &  \ell
\end{pmatrix}
~,\quad
\mathbf{S'}^{\{\alpha'\beta'\}}=
\begin{pmatrix}
\chi_s     &  \chi_f    \\
   &  \nu^c    \\
\end{pmatrix}~.
\label{AllF}
\end{align}
The SM part follows the common Weyl spinor notation, with the additional RH neutrino $\nu^c$, replacing $\psi_0$ in eq. \eqref{fermionS}. The dark fermions from the extension are categorized into two types, $\chi$ and $\psi$, based on whether they are charged under the SM or not. $\chi$ fermions are SM-singlets and can serve as dark quarks of the dark QCD sector. Their subscripts reflect their diverse representations under the $SU(N)_D$ gauge group. On the other hand, $\psi$ fermions are heavy exotic fermions carrying SM charges, with subscripts indicating whether they are charged under the SM $SU(3)_C$ or $SU(2)_W$. Each type of new dark fermion comes with three generations, just like the SM fermions. The potential flavor structure will be discussed in section \ref{sec:Yukawa}. This extended fermion sector is the central ingredient of this study.

\subsection{The gauge sector from the maximal anomaly-free symmetry}\label{sec:gauge}

Next, we turn to the gauge sector. So far, we have the extended gauge group $SU(N+5)$ as well as the gauged global symmetry $SU(N+1)'$. For the fermion content in eq. \eqref{AllF}, there is also an additional anomaly-free $U(1)''$ symmetry, under which the fermions carry the charges
\begin{align}
Q''(\mathbf{A}) =N+1
~,\quad
Q''(\overline{\mathbf{F}}) =-(N+3)
~,\quad
Q''(\mathbf{S'}) = N+5~.
\end{align}
Taking $N=0$, we recover the $SU(5)$ GUT with a RH neutrino $\nu^c$, and this $U(1)$ symmetry, with $\{\mathbf{10}_F,\overline{\mathbf{5}}_F,\nu^c\}$ carrying charges $\{1,-3,5\}$, is known as the $X$ charge and is related to $B-L$ through
\begin{align}\label{B-L}
B-L=\frac{1}{5}\left(4Y+X\right)~.  
\end{align}
Therefore, this additional $U(1)$ symmetry is closely related to $B-L$, as will be further illustrated in section \ref{sec:U(1)}.

As is well known from the swampland conjectures, global symmetries should either be gauged or be broken at high energies \cite{Banks:2010zn}. In this study, we therefore consider that all the anomaly-free global symmetries are gauged in the far UV, resulting in the maximal gauge group
\begin{align}\label{AllG}
SU(N+5) \times SU(N+1)' \times U(1)''~   
\end{align}
which should ultimately be broken down to the SM plus a dark QCD $SU(N)_D$. The fermion content in eq. \eqref{AllF} transforms under the UV gauge group as
\begin{align}\label{newfermion'} 
\mathbf{A}=(A,1)_{N+1}\quad\oplus\quad
\overline{\mathbf{F}}= (\bar{F},\bar{F}')_{-(N+3)}\quad\oplus\quad
\mathbf{S'}=(1,S')_{N+5}~.
\end{align}

Such a maximal gauge group should actually be viewed as an extension of the gauge group $SU(5)\times U(1)_X$ with RH neutrinos, rather than merely the conventional Georgi-Glashow $SU(5)$. In this study, we take this maximal gauge sector in the UV as the starting point and use it as guidance for constructing the whole theory.

Although the resulting gauge and fermion sector might look somewhat complicated at first glance, this system has already been studied before due to different motivations. The nontrivial anomaly cancellation condition has already been discussed and studied for a long time \cite{Banks:1980sc, BahaBalantekin:1980tup}, with a specific interest in strongly coupled chiral gauge theories \cite{Dimopoulos:1980hn,Bars:1981se} which can feature light composite fermions, particularly at the time in the context of preon models \cite{Harari:1979gi,Shupe:1979fv}.

The specific case of $SU(6)\times SU(2)'$ has been considered as a gauge group candidate for GUTs \cite{Rizos:1997pe,Abe:2001qn,Bernard:2012vf} and later for models unifying the SM with dark matter \cite{Barr:2011cz,Barr:2013tea,Barr:2015lya,Chung:2024nnj}. More recently, the possibility of realizing strongly coupled chiral gauge theories in this framework has been revisited \cite{Ryttov:2019evz}, especially as a common setup for symmetric mass generation \cite{Tong:2021phe,Wang:2022ucy}. Some of these perspectives will be discussed again in the following sections.

%\newpage

\section{The Model}\label{sec:Model}

The main idea of this study is based on the extension of the SM fermion content shown in eq. \eqref{AllF}, as well as the gauge sector in eq. \eqref{AllG}, which naturally introduce an $SU(N)_D$ dark sector with dark quark candidates in various representations and three generations. However, to further study the low energy phenomenology, we need to introduce the scalar sector that can trigger the symmetry breaking of the gauge sector and provide the relevant Yukawa interactions and fermion mass terms. This part is rather arbitrary and lacks clear guidance, with the only known scalar being the electroweak Higgs doublet, which, however, is not enough. At the same time, there are many possible choices for the scalar sector, potentially leading to diverse phenomenological properties. In this study, we adopt a minimal approach, introducing only the scalars necessary to achieve the required symmetry breaking and to generate neutrino masses through the Type-I seesaw mechanism \cite{Minkowski:1977sc,Gell-Mann:1979vob,Yanagida:1979as,Glashow:1979nm,Mohapatra:1979ia}. We will analyze the resulting symmetry structure under this setup.

\subsection{The scalar sector and symmetry breaking chain}\label{sec:ScalarSB}

The minimal scalar sector should at least fulfill two goals: 

\renewcommand{\labelenumi}{(\arabic{enumi})} 

\begin{enumerate}

    \item Break the gauge group from the maximal $SU(N+5) \times SU(N+1)' \times U(1)''$ to the SM plus dark color, $SU(N)_D \times SU(3)_C \times U(1)_{EM}$.
    \item Give masses to all SM fermions, including neutrinos, as well as to the exotic fermions carrying SM charges.\footnote{In principle, the dark quarks can be massless without violating any phenomenological constraints, but we will see later that they can acquire masses from the same vacuum as the exotic fermions.}
    
\end{enumerate}
In this subsection, we will introduce the scalars one by one following the breaking chain. The Yukawa interactions and mass terms are addressed in the next subsection.

There are many possible breaking chains, and each can require a very different set of scalars. As most of them only appear at a high scale with negligible effects on the low energy phenomenology, the choice of the breaking chain should not significantly affect our results and conclusions. We therefore focus on a specific breaking chain as follows
\begin{align}
SU(N+5) &\times SU(N+1)' \times U(1)''~\\
\xrightarrow{~\Lambda_{\rm DGUT}~} ~ 
&SU(5) \times SU(N) \times U(1)_N \times SU(N+1)'\times U(1)''\\
\xrightarrow{~\Lambda_{\rm GUT}~} ~ 
&SU(3)_C \times SU(2)_W \times U(1)_Y \times SU(N) \times U(1)_N\times SU(N+1)'\times U(1)''\\
\xrightarrow{~\Lambda_{\rm Seesaw}~} ~
&SU(3)_C \times SU(2)_W \times U(1)_Y \times SU(N) \times U(1)_N\times SU(N)'\times U(1)''_N\label{leftover}\\
\xrightarrow{~\Lambda_{\rm Diagonal}~} ~
&SU(3)_C \times SU(2)_W \times U(1)_Y \times  SU(N)_D \times U(1)_D~\\
\xrightarrow{~\Lambda_{\rm EW}~} ~
&SU(3)_C \times U(1)_{EM} \times SU(N)_D \times U(1)_D~\\
(\xrightarrow{~\Lambda'~} ~
&SU(3)_C \times U(1)_{EM}  \times SU(N)_D)~,
\end{align}
where the required scalar content at different scales will be discussed step by step. Although we mention GUTs frequently, we will not pursue gauge coupling unification but instead focus on the matter content and symmetry structure in this study. ~\\

\paragraph{(1) DGUT breaking : $SU(N+5)\to SU(N) \times SU(5) \times U(1)_N$}~\\

We start with the separation of $SU(N)$ from $SU(N+5)$ as the first breaking and call it DGUT, standing for Dark Grand Unified Theory, since at this stage (part of) the dark color gauge group is unified with the $SU(5)$ GUT. The breaking is chosen at the first step because $SU(N+5)$ can easily become strongly coupled and confine at a rather high energy if there is insufficient matter content.

Following the conventional GUT breaking setup, we introduce a complex scalar $\mathbf{\Gamma}$ transforming as an adjoint representation of $SU(N+5)$, which preserves the rank, and as a singlet under $SU(N+1)' \times U(1)''$. Once it acquires a non-zero vacuum expectation value (VEV) given by
\begin{align}\label{GammaVEV}
\langle \mathbf{\Gamma} \rangle = \langle \Gamma_0 \rangle = \Lambda_{\rm DGUT}\times T_N~,~\text{where}~
T_N=
\begin{pmatrix}
5\,\mathbb{I}_{N\times N}  &  0    \\
0  &  -N\,\mathbb{I}_{5\times 5}    \\
\end{pmatrix}~,
\end{align} 
it breaks $SU(N+5)\to SU(N) \times SU(5) \times U(1)_N$, where $T_N$ is the generator of the $U(1)_N$ symmetry. The scalar $\mathbf{\Gamma}$ decomposes into a set of complex scalars $\Gamma_i$ with quantum numbers
\begin{align}\label{Gamma}
\mathbf{\Gamma} = \text{Adj} \to~ 
\Gamma_N=(\text{Adj},1)_0 ~\oplus~ \Gamma_5=(1,\text{Adj})_0 ~\oplus~ \Gamma_0=(1,1)_0 ~\oplus~ \Gamma_\ell=(N,\bar{5})_{N-5}
\end{align}
under the unbroken $SU(N) \times SU(5) \times U(1)N$. The component $\Gamma_0$ acquires a VEV, while $\Gamma_\ell$ becomes the longitudinal component of the massive gauge bosons. Under this breaking, a total of $10N$ gauge bosons, which couple the dark fermions to the SM fermions, become massive at the scale $\Lambda_{\rm DGUT}$, along with the rest of scalars $\Gamma_i$. In certain models \cite{Murgui:2021eqf}, these gauge bosons can lead to dark matter decay and therefore need to be sufficiently heavy, providing another reason for this breaking scale to be high.

\paragraph{(2) GUT breaking : $SU(5)\to SU(3)_C \times SU(2)_W \times U(1)_Y$}~\\

Next, we have our conventional GUT breaking, which can be realized by the already introduced scalar $\Gamma_5\subset \mathbf{\Gamma}$. Once $\Gamma_5$ acquires a VEV give by
\begin{align}\label{Gamma5V}
\langle \Gamma_5 \rangle =  \Lambda_{\rm GUT}\times T_Y~,~\text{where}~
T_Y=
\begin{pmatrix}
-\frac{1}{3}\,\mathbb{I}_{3\times 3}  &  0    \\
0  &  \frac{1}{2}\,\mathbb{I}_{2\times 2}    \\
\end{pmatrix}~,
\end{align}
where $T_Y$ is the generator of the $U(1)_Y$ hypercharge, the $SU(5)$ GUT is broken down to the SM gauge group $SU(3)_C \times SU(2)_W \times U(1)_Y$. The $12$ massive gauge bosons that can mediate the proton decay are introduced at this scale. From the latest proton decay constraints \cite{Super-Kamiokande:2020wjk}, $\Lambda_{\rm GUT}\gtrsim 10^{16}$ GeV is required. 

As both $\Gamma_0$ and $\Gamma_5$ arise from the same multiplet $\mathbf{\Gamma}$, in the absence of a splitting mechanism in the scalar potential or unusual quartic couplings, we expect the two scales to be roughly the same, $\Lambda_{\rm DGUT}\sim \Lambda_{\rm GUT}$, and the ordering is chosen merely for convenience.~\\

\paragraph{(3) Seesaw and lepton number breaking: $SU(N+1)'\times U(1)''\to SU(N)'\times U(1)''_N$}~\\

The third step is a rather non-general setup because the origin of neutrino masses is not yet determined. However, with the presence of RH neutrinos, it is straightforward to include the Type-I seesaw mechanism \cite{Minkowski:1977sc,Gell-Mann:1979vob,Yanagida:1979as,Glashow:1979nm,Mohapatra:1979ia} by introducing the RH neutrino masses at the seesaw scale $\Lambda_{\rm Seesaw}$. This can be achieved by introducing an additional scalar
\begin{align}
\mathbf{\Delta}_{\{\alpha'\beta'\gamma'\delta'\}}~~\text{with}~~
{\cal L}_{\rm Seesaw} = - \mathbf{Y}_\Delta \mathbf{\Delta}_{\{\alpha'\beta'\gamma'\delta'\}}
\mathbf{S'}^{\{\alpha'\beta'}\mathbf{S'}^{\gamma'\delta'\}} +\text{h.c.}~,
\end{align}
where $\alpha'=1,2,...N+1$ is an $SU(N+1)'$ index and $\mathbf{Y}_\Delta$ is a $3\times 3$ Yukawa coupling matrix. The RH neutrino masses are generated once the scalar $\mathbf{\Delta}$ acquires a VEV $v_R$ at the entry with $\alpha'=\beta'=\gamma'=\delta'=N+1$ given by
\begin{align}\label{DeltaV}
\langle \mathbf{\Delta} \rangle = \langle \Delta_0 \rangle = \frac{v_R}{\sqrt{2}}~\implies \mathbf{M}_R=\mathbf{Y}_\Delta \times \frac{v_R}{\sqrt{2}}~,
\end{align}
where $\mathbf{M}_R$ is a $3\times 3$ mass matrix. Taking the eigenvalues of $\mathbf{Y}_\Delta$ to be of $\mathcal{O}(1)$, the typical seesaw mechanism requires $v_R\sim\Lambda_{\rm Seesaw}\sim 10^{10-14}$ GeV, depending on the details of the Dirac neutrino masses, which will be discussed in the next section.

The VEV of the scalar $\mathbf{\Delta}$ in eq.~\eqref{DeltaV} not only introduces the required RH neutrino mass terms for the Type-I seesaw mechanism but also breaks the corresponding gauge symmetries. With the extension of the gauge sector, the RH neutrino $\nu^c$ is part of the multiplet $\mathbf{S'}$, which transforms under $SU(N+1)'$ and also carries a $U(1)''$ charge of $N+5$. Therefore, the scalar breaks the $SU(N+1)'\times U(1)''$ down to a subgroup $SU(N)'\times U(1)''_N$ with a total of $2N+1$ broken generators.

The breaking contains both non-Abelian and Abelian parts. The non-Abelian part is rather simple, where the $2N$ off-diagonal generators are broken and the corresponding gauge bosons, which couple the RH neutrinos to the dark quarks $\chi_s$ and $\chi_f$, become massive with masses of $\mathcal{O}(\Lambda_{\rm Seesaw})$.

However, for the abelian part, it is rather nontrivial. In the SM, we know that the RH neutrino masses break the lepton number symmetry, which, under the GUT framework, is given by a combination of the $X$ charge and $Y$ hypercharge as shown in eq.~\eqref{B-L}. In our extended framework, this symmetry takes a more complicated form. To keep track of the $U(1)$ symmetry breaking part, we first extract the $U(1)$ subgroup
\begin{align}\label{U(1)'N}
U(1)'_N\subset SU(N+1)'~~\text{with the generator}~~
T'_N=
\begin{pmatrix}
\mathbb{I}_{N\times N}  &  0    \\
0  &  -N\,\mathbb{I}_{1\times 1}    \\
\end{pmatrix}~
\end{align} 
up to a normalization factor. The $U(1)$ symmetry breaking that corresponds to the lepton number symmetry breaking is then given by
\begin{align}\label{U(1)''N}
U(1)'_N\times U(1)''\to U(1)''_N
\end{align} 
where the generator
\begin{align}\label{T''N}
T''_N=(N+5)\times T'_N+2N\times T''~.
\end{align} 
The charges of fermions under the $U(1)''_N$ are given by
\begin{align}
&Q''_N(\mathbf{A}) =2N(N+1)
~,\quad
Q''_N(\overline{\mathbf{F}}) =
\begin{pmatrix}
-(2N+5)(N+1)~~  &  -N(N+1)  \\ 
-(2N+5)(N+1)~~  &  -N(N+1)  \\ 
-(2N+5)(N+1)~~  &  -N(N+1)  \\
\end{pmatrix}~,\nonumber\\
&
Q''_N(\mathbf{S'}) = Q''_N
\begin{pmatrix}
\chi_s     &  \chi_f    \\
   &  \nu^c    \\
\end{pmatrix}
= 
\begin{pmatrix}
2(N+1)(N+5)~~  &   (N+1)(N+5)    \\
   &   0       \\
\end{pmatrix}~,
\end{align}
where the RH neutrino is chargeless. The breaking introduces a heavy neutral boson, denoted as $Z_R$. The boson acquires a mass at the seesaw scale and is therefore not of phenomenological interest at low energies. We therefore defer the detailed study of $Z_R$ to Appendix \ref{App:U(1)SB}. In the main text, further analysis of the $U(1)$ symmetries and their relation to lepton number $L$ will be presented in section~\ref{sec:U(1)}.~\\

\paragraph{Brief summary of the first three high-scale symmetry breaking scales}~\\

Yet the three breaking scales we discuss are expected to be well above the accessible scale, either due to existing constraints, such as those from proton decay, or due to the requirement of reproducing the tiny neutrino masses. Therefore, the details, such as the exact values and ordering of these scales, are largely irrelevant to the phenomenology of interest, and we will only focus on the resulting symmetry structure. It is nevertheless worth mentioning that these heavy particles, although not directly accessible experimentally, might play an important role in the early Universe, for example, in the production of dark matter through mechanisms such as Leptogenesis \cite{Fukugita:1986hr}. We leave a detailed study of these possibilities for future work and focus on the other BSM particles that can be light and affect low energy phenomenology.

%\newpage

\paragraph{(4) Diagonal breaking : $SU(N)\times SU(N)'\times U(1)_N\times U(1)''_N\to SU(N)_D\times U(1)_D$}~\\

Next, we move on to the most essential step of gauge symmetry breaking in the model, $SU(N)\times SU(N)'\to SU(N)_D$, which we denote as the diagonal breaking. There are many scalars that can realize this breaking. However, as we mentioned in section \ref{sec:Idea}, the reason for having the dark color gauge group $SU(N)_D$ arise as the diagonal subgroup of $SU(N)\times SU(N)'$ is to avoid the exotic fermions $\psi$ from being massless. Therefore, we would like to have a scalar which not only realizes the gauge symmetry breaking but also generates masses for the exotic fermions. The desired scalar then transforms as
\begin{align}
{\mathbf{\Phi}^\alpha}_{\alpha'}= (F,\bar{F}')_{-2}=
\begin{pmatrix}
~~\Phi_0=\phi_0\oplus\phi_{\rm adj}~ & \phi_\nu \\ \phi_d & T_d \\ \phi_\ell &  H_d
\end{pmatrix}
\label{Phi}~
\end{align}
under $SU(N+5) \times SU(N+1)' \times U(1)''$. We also show its decomposition after the symmetry breaking in matrix form. These scalars transform as
\begin{align}\label{newscalar} 
&\Phi_0=\phi_0\oplus\phi_{\rm adj}= (1,1,1)_0 \oplus ({\rm Adj},1,1)_0~,~\phi_d=(\bar{N},3,1)_{-\frac{1}{3}}~,~\phi_\ell =(\bar{N},1,2)_{\frac{1}{2}}~,\nonumber\\
&\phi_\nu= (N,1,1)_0~,~ T_d=(1,3,1)_{-\frac{1}{3}}~,~ H_d =(1,1,2)_{\frac{1}{2}}
\end{align}
under the gauge group $SU(N)_D \times SU(3)_C \times SU(2)_W \times U(1)_Y$.

One can immediately find that one of the scalars has the same quantum number as the SM Higgs. We therefore denote it as $H_d$, following the Yukawa coupling it introduces, as will be discussed in the next subsection. Together with the color triplet scalar $T_d$, they form the $\overline{\mathbf{5}}_H$ commonly appearing in $SU(5)$ GUTs. Therefore, the scalar $\mathbf{\Phi}$ can also be viewed as an extension of the Higgs doublet. Three dark-colored (anti-)fundamental scalars, $\phi_\nu$, $\phi_d$, and $\phi_\ell$, are labeled by the SM fermions with which they couple through the Yukawa coupling terms, which will also be illustrated in the next subsection. They play an important role as portals, which will be the main focus in section \ref{sec:Scalar}.

In this section, we focus on the complex scalar $\phi_0$, which acquires a VEV given by
\begin{align}\label{PhiVEV}
\langle \mathbf{\Phi} \rangle = \langle {\Phi_0} \rangle =\langle \phi_0 \rangle = \frac{f}{\sqrt{2}}
\begin{pmatrix}
\mathbb{I}_{4\times 4}  &  0    \\
0  &  0    \\
\end{pmatrix}~.
\end{align} 
The $\phi_{\rm adj}$ components then become the Goldstone bosons and are eaten by the gauge bosons of the broken $SU(N)_A$, resulting in $N^2-1$ heavy dark colorons with a mass of $\mathcal{O}(f)$, where we borrow the term from the Topcolor model \cite{Hill:1991at,Chivukula:1996yr} that shares a similar gauge structure.

The VEV $f$ sets the typical scale of $\Lambda_{\rm Diagonal}$, at which $\Phi_0$ and the dark colorons are expected to be located. However, it is not a generic scale for all the scalars from the multiplet $\mathbf{\Phi}$ shown in eq.~\eqref{newscalar}. For example, we already know that the electroweak doublet $H_d$ should be around $\Lambda_{\rm EW}$, while the color triplet $T_d$, which can mediate proton decay, should be at least as heavy as $\Lambda_{\rm GUT}$. Already within the $SU(5)$ GUT, such a scalar multiplet suffers from the well-known doublet-triplet splitting problem, and there is no clear guidance for determining the spectrum of $\phi_\nu$, $\phi_d$, and $\phi_\ell$.

In this study, we therefore consider two possible assumptions. One is simply to assume that the masses of the new dark-colored scalars arise from $\langle \Phi_0\rangle$, giving
\begin{align}\label{Mphi1}
\text{Assumption (1) : }
M_{\phi_d}\sim M_{\phi_\ell} \sim M_{\phi_\nu} \sim \Lambda_{\rm Diagonal} \sim f ~.
\end{align} 
On the other hand, it is also possible that the $SU(N+1)'$ symmetry, though broken at this stage, still remains relevant for the scalar potential, resulting in
\begin{align}\label{Mphi2}
\text{Assumption (2) : }
M_{\phi_d}\sim \Lambda_{\rm GUT}~,\quad
M_{\phi_\ell} \sim ~ \Lambda_{\rm EW}~,\quad
M_{\phi_\nu}  \sim \Lambda_{\rm Diagonal} \sim f ~.
\end{align} 
The spectrum of these scalars has an important impact on the low energy phenomenology, which will be discussed in section \ref{sec:Scalar}.

In this study, we realize the symmetry breaking through the VEV of an elementary scalar. However, it is also possible to realize it dynamically through strong dynamics of either the $SU(N)$ or $SU(N)'$ gauge interaction, which is chiral at this stage and may trigger a self-breaking mechanism \cite{Amati:1980wx,Gusynin:1982kp}, with similar realizations studied in \cite{Martin:1992aq,Martin:1992mj,Chung:2023iwj}. Whether such a phase is indeed realized is still under investigation \cite{Csaki:2021xhi,Bolognesi:2021jzs,Smith:2021vbf,Karasik:2022gve,Li:2025tvu,Li:2026ayh}. Nevertheless, condensate candidates with the same quantum numbers as $\Phi_0$ can be readily constructed, such as $\langle \chi_a \psi_f^c \rangle$ under $SU(N)$ and $\langle \chi_s \psi_f^c \rangle$ under $SU(N)'$. However, if the symmetry breaking is realized in this way, the scalar arising as a composite state of $SU(N)$ or $SU(N)'$ will not form a complete multiplet under the full gauge group $SU(N+5) \times SU(N+1)' \times U(1)''$, resulting in a reduced scalar sector. We leave such a realization for future study. We leave a detailed study of such a dynamical realization for future work.

Back to the symmetry breaking, notice that $\phi_0$ not only realizes the diagonal breaking $SU(N)\times SU(N)'\to SU(N)_D$ for the non-abelian groups but also induces an additional $U(1)_N\times U(1)''_N\to U(1)_D$ symmetry breaking, since $\phi_0$ carries nonzero charges under $U(1)_N$ and $U(1)''_N$ with $Q_N=5$ and $Q''_N=-5N-5$. The breaking introduces a heavy neutral gauge boson $Z'$ with a mass of $\mathcal{O}(f)$. Its detailed properties are presented in Appendix \ref{App:U(1)SB}. If the scale $f$ is sufficiently light, the $Z'$ can lead to relevant phenomenology and connect the dark sector with the visible sector, as will be discussed in section \ref{sec:Z'}.

The remaining unbroken $U(1)_D$ symmetry is given by
\begin{align}\label{U(1)D}
c~T_D=(N+1)\times T_N + T''_N=(N+1)\times T_N + (N+5)\times T'_N+2N\times T''~,
\end{align} 
where $c$ is a normalization factor. Taking $c=(N+1)(N+5)N$, we obtain the charges of the fermions, written in matrix form to match eq.~\eqref{AllF}, as
\begin{align}\label{Dcharge} 
D(\mathbf{A}) =\frac{1}{N}
\begin{pmatrix}
2      &   1   &  1 \\
   &   0      &  0      \\
   &       &  0   \\
\end{pmatrix}
~,\quad
D(\overline{\mathbf{F}}) =\frac{1}{N}
\begin{pmatrix}
-2  &  -1\\ 
-1  &  0   \\ 
-1  &  0  \\
\end{pmatrix}
~,\quad
D(\mathbf{S'}) =\frac{1}{N}
\begin{pmatrix}
2      &   1    \\
   &   0       \\
\end{pmatrix}~.
\end{align}
We can easily find that only the dark fermions carry a nonzero charge. The charges thus reflect the number of dark fermions and can serve as the \textbf{Dark Baryon Number $D$}. Due to our interest in a strongly coupled dark sector, we choose the normalization factor such that the fundamental dark quark $\chi_f$ carries charge $1/N$ such that the typical dark baryons of $N$ fundamental dark quarks carry dark baryon number $D=1$.

The same symmetry and charge have also been identified in similar frameworks \cite{Barr:2011cz,Barr:2013tea,Barr:2015lya}. However, in our model, it is not merely an accidental global symmetry but an unbroken gauge symmetry, featuring a massless gauge boson, the dark photon $A'$ \cite{Holdom:1985ag,Okun:1982xi}. Depending on the model, the existence of a massless dark photon can be either beneficial or dangerous. We therefore leave the breaking scenario of the $U(1)_D$ open, with a characteristic scale $\Lambda'$ ranging all the way from $\Lambda'=0$, featuring a massless dark photon, to $\Lambda'>\Lambda_{\rm DGUT}$, where the dark photon is decoupled from all relevant physics. This can be achieved, for example, by breaking $U(1)''$ by a charged scalar in the far UV. In the following discussion, we assume that the dark photon is massless or sufficiently light that it cannot be integrated out.~\\

\paragraph{(5) Electroweak symmetry breaking : $SU(2)_W \times U(1)_Y\to U(1)_{EM}$}~\\

The last step is electroweak symmetry breaking (EWSB), which can be realized by the electroweak doublet $H_d\subset \mathbf{\Phi}$ that we have already introduced. However, breaking the symmetry is not enough, as the scalar should also be able to reproduce the SM fermion masses. From $H_d\subset \mathbf{\Phi}$, we can already generate the masses for down-type quarks and charged leptons, as well as Dirac neutrino masses, which will be illustrated in the next section. However, to generate the masses of up-type quarks, an additional Higgs doublet from another scalar multiplet is required. We therefore introduce a scalar 
\begin{align}
\mathbf{H}^{[\alpha\beta....]}= (A_{N+1},1)_{-2(N+1)}~,
\label{H}~
\end{align}
transforming in the rank-$(N+1)$ antisymmetric representation of $SU(N+5)$.

To understand this scalar, we focus on its quantum numbers under $SU(N+5)$. After the breaking $SU(N+5)\to SU(N) \times SU(5)$, the scalar decomposes into five terms as
\begin{align}\label{Hall}
\mathbf{H}^{[\alpha\beta....]}~=~ 
H^{[a,ijk...]} ~\oplus~ H^{[ab,ijk...]} ~\oplus~ H^{[abc,ijk...]}~\oplus~ H^{[abcd,ijk...]}~\oplus~ H^{[abcde,ijk...]}
\end{align}
according to the different number of indices $n$ transforming under $SU(5)$ (we use $a,b,c,d,e$) or $SU(N)$ (we use {$i,j,k....$}), where $n=1,2,3,4,5$. For the $SU(5)$ part, we directly denote the corresponding representations by their dimensions. For the $SU(N)$ part, since the scalars possesses $N+1-n$ totally antisymmetric upper indices, they transform in representations that are isomorphic to those with $n-1$ lower indices. Therefore, we can rewrite the five components as
\begin{align}\label{Hall}
\mathbf{H}^{[\alpha\beta....]} ~=~ 
\mathbf{5}_H ~\oplus~ \mathbf{10}_{H,i} ~\oplus~ \overline{\mathbf{10}}_{H,ij}~\oplus~ \overline{\mathbf{5}}_{H,ijk}~\oplus~ \mathbf{1}_{H,ijk\ell}~.
\end{align}

Starting from the first scalar, the fiveplet $\mathbf{5}_H$, which can be further decomposed under the SM gauge group as
\begin{align}
\mathbf{5}_H ~\to~ T_u=(1,3,1)_{-\frac{1}{3}} ~\oplus~ H_u =(1,1,2)_{\frac{1}{2}}~,
\end{align}
which contains a color triplet $T_u$, which should be as heavy as $\Lambda_{\rm GUT}$ to avoid rapid proton decay and a Higgs doublet $H_u$, which is required to generate the up-type quark masses.

With $H_u \subset \mathbf{H}$ and $H_d \subset \mathbf{\Phi}$, we get a Type-II Two-Higgs-Doublet Model (2HDM) \cite{Branco:2011iw}, which satisfies natural flavor conservation \cite{Glashow:1976nt, Paschos:1976ay} due to the symmetry structure of our extended gauge group. The observed electroweak scale is then generated by
\begin{align}
\langle H_u\rangle=\frac{v_u}{\sqrt{2}}~,\quad
\langle H_d\rangle=\frac{v_d}{\sqrt{2}}~\implies~
v_{\rm EW} =\sqrt{v_u^2+v_d^2}=246~{\rm GeV}~
\end{align}
with $\tan\beta \equiv v_u/v_d$. We assume that the extended Higgs sector resides within the decoupling limit, safely satisfying current constraints from Higgs measurements, direct searches and flavor observables \cite{Misiak:2017bgg,HFLAV:2019otj,Kling:2020hmi,Li:2024kpd}. Therefore, the phenomenology of the Type-II 2HDM will not be discussed in the following study.

The remaining scalars transform under the $SU(N)D$ gauge group, including the anti-fundamental and rank-2,3,4 antisymmetric representations. We present their decompositions under the SM gauge group, following the matrix notation in eq.~\eqref{SU(5)F}, as
\begin{align}
\mathbf{10}_{H,i}=
\begin{pmatrix}
\tilde{u}^c_{i} &  \tilde{q}_{i}  \\
    &  \tilde{e}^c_{i}  \\
\end{pmatrix}
,~
\overline{\mathbf{10}}_{H,ij}=
\begin{pmatrix}
\phi_{u,ij}~ &  \phi_{q,ij}  \\
    &  \phi_{e,ij}  \\
\end{pmatrix}
,~
\overline{\mathbf{5}}_{H,ijk}=
\begin{pmatrix}
\tilde{d}_{ijk}\, \\ \tilde{\ell}^c_{ijk}\,
\end{pmatrix}
,~
\mathbf{1}_{H,ijk\ell}=\sigma_{ijk\ell}
\label{H_decomposition}
\end{align}
where the $SU(N)D$ indices are kept explicit to emphasize their dark color charges. Regarding the notation, since the scalars in $\mathbf{10}_{H,i}$ and $\overline{\mathbf{5}}_{H,ijk}$ share the same SM gauge quantum numbers as the SM fermions, we adopt the sfermion notation used in supersymmetry. For the SM singlet $\mathbf{1}_{H}$, we simply denote it by $\sigma$.

Among the four dark-colored scalar multiplets, $\overline{\mathbf{10}}_{H,ij}$ is of particular interest, which can be further decomposed into three scalars, $\phi_u$, $\phi_q$, and $\phi_e$, transforming as 
\begin{align}
\overline{\mathbf{10}}_{H,ij} ~\to~ \phi_u=(\bar{A},3,1)_{\frac{2}{3}} ~\oplus~ \phi_q =(\bar{A},\bar{3},\bar{2})_{\frac{1}{2}}~\oplus~ \phi_e =(\bar{A},1,1)_{-1}~,
\end{align}
under $SU(4)_D\times SU(3)_C \times SU(2)_W \times U(1)_Y$. These scalars couple the dark quark $\chi_a$ to the SM fermions $u,q,e$ through the Yukawa coupling terms, which will be illustrated in the next subsection. Therefore, we denote them by the SM fermions with which they couple, similar to the previously introduced scalars $\phi_{d,\ell,\nu} \subset \mathbf{\Phi}$.

Similar to the scalar multiplet $\mathbf{\Phi}$, there is no clear guidance for the spectrum of scalars from $\mathbf{H}$ as doublet-triplet splitting issue also arise in $\mathbf{5}_H\subset \mathbf{H}$. Without a detail study of scalar potential and a concrete solution to splitting problem, it is hard to make any claim on the scalar spectrum. We leave the study to future and here we only make assumptions matching the previous treatment. Again, for one we assume that the masses of the new dark-colored scalars arise from $\langle \Phi_0\rangle$, giving
\begin{align}\label{Mphi1*}
\text{Assumption (1) : }
M_{\phi_u}\sim M_{\phi_q}\sim M_{\phi_e}  \sim \Lambda_{\rm Diagonal} \sim f ~,
\end{align} 
which can originate from the potential like $\mathbf{H}^\dagger\mathbf{\Phi}^\dagger\mathbf{\Phi}\mathbf{H}$. The other, assuming the doublet-triplet splitting also apply to the dark-colored scalars, which is the case for solutions like sliding singlet \cite{Witten:1981kv,Nanopoulos:1982wk,Dimopoulos:1982af,Ibanez:1982fr,Csaki:2024ywk} and related to the potential like $\mathbf{H}^\dagger\mathbf{\Gamma}\mathbf{H}$, will then gives
\begin{align}\label{Mphi2*}
\text{Assumption (2) : }
M_{\phi_u}\sim M_{\phi_q}\sim \Lambda_{\rm GUT}~,\quad
M_{\phi_\ell} \sim ~ \Lambda_{\rm EW}~,
\end{align} 
where the colored scalars are heavy while the colorless scalars can remain light. The consequence of these assumptions and the impact on the low energy phenomenology will be further discussed in section \ref{sec:Scalar}.

%\newpage

\paragraph{Summary of the scalar sector and symmetry breaking}~\\

To sum up, the complete scalar content is given below, together with their quantum numbers under $SU(N+5)\times SU(N+1)' \times U(1)''$ and their respective roles:
\begin{itemize}
\item ${\mathbf{\Gamma}^\alpha}_{\beta} = ({\rm Adj},1)_0$ 
realizes the DGUT and GUT symmetry breakings.
\item $\mathbf{\Delta}_{\{\alpha'\beta'\gamma'\delta'\}} = (1,\bar{S}'_4)_{-2(N+5)}$ realizes lepton number breaking and gives masses to the RH neutrinos for the Type-I seesaw mechanism.
\item ${\mathbf{\Phi}^\alpha}_{\alpha'} = (F,\bar{F}')_{-2}$ includes (1) $\Phi_0$, which realizes the diagonal breaking and gives masses to the dark fermions, and (2) $H_d$, which realizes EWSB and gives masses to the SM fermions except the up-type quarks.
\item $\mathbf{H}^{[\alpha\beta....]} = (A_{N+1},1)_{-2(N+1)}$ realizes EWSB and gives masses to the up-type quarks.
\end{itemize}

The four scalar multiplets compose the minimal setup that breaks the maximal gauge group down to the SM plus dark color, with an additional unbroken $U(1)_D$ symmetry, and gives mass to all fermions. Two of the scalars, $\mathbf{\Gamma}$ and $\mathbf{\Delta}$, are expected to be heavy due to their roles. However, the other two scalars, $\mathbf{\Phi}$ and $\mathbf{H}$, which provide essential ingredients for diagonal breaking and electroweak symmetry breaking, can in principle be at experimentally accessible scales and introduce rich phenomenology, and therefore will be our main focus in the following study. In the next subsection, we first discuss the Yukawa couplings and fermion masses generated by these scalars.

%\newpage

\subsection{Yukawa interaction terms and fermion masses}\label{sec:Yukawa}

After the discussion of the scalar content required in the minimal setup, with a focus on its role in the gauge symmetry breaking chain, we now turn our attention to the interactions between the scalar and fermion sectors, i.e., the Yukawa interaction terms. In this subsection, we focus in particular on the mass terms generated by these interactions, some of which have already been mentioned but will be discussed in detail here. Other interactions, such as portal interactions between dark quarks and SM fermions, are left to section \ref{sec:Portal}.

To write down the Yukawa interaction terms that obey the maximal gauge symmetry $SU(N+5) \times SU(N+1)' \times U(1)''$, we first list the complete fermion and scalar contents
\begin{align}
&\text{Fermion:}\quad
\mathbf{A}=(A,1)_{N+1}~,~
\overline{\mathbf{F}}= (\bar{F},\bar{F}')_{-(N+3)}~,~
\mathbf{S'}=(1,S')_{N+5}~\text{ for each generation.}\nonumber\\
&\text{Scalar:}\quad
\mathbf{\Gamma} = ({\rm Adj},1)_0~,~
\mathbf{\Delta} = (1,\bar{S}'_4)_{-2(N+5)}~,~
\mathbf{\Phi} = (F,\bar{F}')_{-2}~,~
\mathbf{H} = (A_{N+1},1)_{-2(N+1)}~.\nonumber
\end{align}
First, we notice that the adjoint scalar $\mathbf{\Gamma}$ does not introduce any Yukawa coupling terms due to the chiral nature of the fermion content. Therefore, only the three remaining scalar multiplets are relevant for the Yukawa interactions. The two scalars $\mathbf{\Delta}$ and $\mathbf{H}$ can couple to $\mathbf{S'}\mathbf{S'}$ and $\mathbf{A}\mathbf{A}$, respectively, which is precisely why they are introduced: to generate the Majorana mass of $\nu^c\subset\mathbf{S'}$ and the up-type quark masses of $u,u^c\subset\mathbf{A}$, respectively. The remaining scalar, $\mathbf{\Phi}$, on the other hand, can couple to both $\mathbf{S'}\overline{\mathbf{F}}$ and $\mathbf{A}\overline{\mathbf{F}}$ and generate the remaining fermion masses, playing an essential role in the model.

%\newpage

\paragraph{Yukawa interaction terms}~\\

We obtain four Yukawa interaction terms that obey the maximal gauge symmetry as
\begin{align}\label{Yukawa}
-{\cal L}_{\rm Yukawa} = 
\mathbf{Y}_\Delta\, \mathbf{\Delta}\mathbf{S'}\mathbf{S'}+
\mathbf{Y}_{\Phi} \,\mathbf{\Phi}\mathbf{S'}\overline{\mathbf{F}}+
\mathbf{Y}_{\Phi^*} \,\mathbf{\Phi}^*\mathbf{A}\overline{\mathbf{F}}+
\mathbf{Y}_H\, \mathbf{H}\mathbf{A}\mathbf{A}
+\text{h.c.}
\end{align}
Since all the fermions come in three generations, the Yukawa couplings $\mathbf{Y}$ are $3\times 3$ matrices that encode nontrivial flavor structures. Next, we expand these terms one by one according to the decompositions discussed in the previous subsection. For convenience, we first ignore the coefficients and flavor indices, showing only representative terms as follows.

\paragraph{The first term:}
\begin{align}\label{DSS}
\mathbf{\Delta}\mathbf{S'}&\mathbf{S'}+\text{h.c.}=~ 
 \mathbf{\Delta}_{\{\alpha'\beta'\gamma'\delta'\}}
\mathbf{S'}^{\{\alpha'\beta'}\mathbf{S'}^{\gamma'\delta'\}} +\text{h.c.}~,\nonumber\\
&=\frac{1}{2}\Delta_0\nu^c\nu^c+\Delta_1\nu^c\chi_f+\Delta_2(\nu^c\chi_s+\frac{1}{2}\chi_f\chi_f)+\Delta_3\chi_f\chi_s+\frac{1}{2}\Delta_4\chi_s\chi_s+\text{h.c.}
\end{align}
where we simply decompose $\mathbf{\Delta}=\Delta_0\oplus\Delta_1\oplus\Delta_2\oplus\Delta_3\oplus\Delta_4$, labeled by the rank of the corresponding representations. In the following discussion, we will focus only on $\Delta_0$. 

\paragraph{The second term:}
\begin{align}\label{PSF}
\mathbf{\Phi}\mathbf{S'}&\overline{\mathbf{F}}+\text{h.c.}=~ 
 {\mathbf{\Phi}^\alpha}_{\alpha'}\mathbf{S'}^{\alpha'\beta'}{\overline{\mathbf{F}}}_{\alpha\beta'}+\text{h.c.}\nonumber\\
&=
\phi_0(\chi_s\chi^c_s+\chi_f\chi^c_f)+
\phi_{\rm adj}(\chi_s\chi^c_{s}+\chi_f\chi^c_{f})+
\phi_d(\chi_s\psi^c_c+\chi_fd^c)+\phi_\ell(\chi_s\psi^w_c+\chi_f\ell)\nonumber\\
&+
{\phi_\nu}(\chi_f(\chi^c_{a}+\chi^c_{s})+\nu^c\chi^c_{f})+
T_d(\chi_f\psi^c_c+\nu^cd^c)+H_d(\chi_f\psi_w^c+\nu^c\ell)
+\text{h.c.}
\end{align}

\paragraph{The third term:}

\begin{align}\label{PAF}
\mathbf{\Phi}^*\mathbf{A}&\overline{\mathbf{F}}+\text{h.c.}=~ 
 {{\mathbf{\Phi}^*}_\alpha}^{\alpha'}\mathbf{A}^{\alpha\beta}{\overline{\mathbf{F}}}_{\beta\alpha'}+\text{h.c.}\nonumber\\
&=
\phi_0^*(\chi_a\chi^c_a+\psi_c\psi_c^c+\psi_w\psi_w^c)+
\phi_{\rm adj}^*(\chi_a\chi^c_a+\psi_c\psi^c_{c}+\psi_w\psi^c_{w})
\nonumber\\
&+
\phi_d^*(\psi_c(\chi^c_a+\chi^c_s)+u^c\psi_c^c+q\psi_w^c)
+\phi_\ell^*(\psi_w(\chi^c_a+\chi^c_s)+q\psi_c^c+e^c\psi_w^c)
\nonumber\\
&+
{\phi_\nu^*}(\chi_a\chi_f^c+\psi_cd^c+\psi_w\ell)+
T_d^*(\psi_c\chi_f^c+u^cd^c+q\ell)+
H_d^*(\psi_w\chi_f^c+qd^c+e^c\ell)
+\text{h.c.}
\end{align}

For the fourth term, we expand it in two steps, following the decomposition of $\mathbf{H}$. First, we separate the $SU(5)$ and $SU(N)$ indices and introduce the fiveplet $\psi^i=(\psi_c^i,\psi_w^i)$. Then, we further decompose these terms into the states below the GUT scale as follows.

\paragraph{The fourth term:}

\begin{align}\label{HAA}
\mathbf{H}\mathbf{A}\mathbf{A}\,+&\,\text{h.c.}=~ 
\mathbf{H}^{[\rho\sigma....}
\mathbf{A}^{\alpha\beta}\mathbf{A}^{\gamma\delta]} +\text{h.c.}~,\nonumber\\
=~&\mathbf{5}_H\mathbf{10}_F\mathbf{10}_F + 
\mathbf{10}_{H,i}\mathbf{10}_F\psi^i + 
\overline{\mathbf{10}}_{H,ij}\mathbf{10}_F\chi_a^{[ij]}\nonumber\\
&+\overline{\mathbf{10}}_{H,ij}\psi^i\psi^j+ \overline{\mathbf{5}}_{H,ijk}\chi_a^{[ij]}\psi^k+
\mathbf{1}_{H,ijk\ell}\chi_a^{[ij]}\chi_a^{[k\ell]}+\text{h.c.}\nonumber\\
=~&[H_uqu^c+T_u(qq+u^ce^c)]\nonumber\\
&+
[(\tilde{e}u^c+\tilde{u}e^c+\tilde{q}q)\psi_c+(\tilde{q}u^c+\tilde{u}q)\psi_w]+
[(\phi_uu^c+\phi_qq+\phi_ee^c)\chi_a] \nonumber\\
&+
[\phi_u\psi_c\psi_c+\phi_q\psi_c\psi_w+\phi_e\psi_w\psi_w]+
[\tilde{d}\chi_a\psi_c+\tilde{\ell}\chi_a\psi_w]+
\sigma\chi_a\chi_a+\text{h.c.,}
\end{align}
where the terms describe interactions among SM fermions, between SM and dark fermions, and among dark fermions, respectively.

\newpage

Next, to discuss the fermion masses, we focus on the terms involving scalars that acquire VEVs, namely $\Delta_0$, $\phi_0$, $H_d$, and $H_u$, which appear in either the first or the last terms of the equations above. We first discuss the SM fermion masses including neutrinos, followed by the dark fermion masses.~\\

\paragraph{The SM fermion and neutrino masses}~\\

Start with the SM charged fermions, whose masses are generated from the VEVs of the Higgs doublets. From the terms with $H_d$ and $H_u$ in eq.~\eqref{PAF} and eq.~\eqref{HAA}, we obtain
\begin{align}\label{SMMass}
-{\cal L}_{\rm Yukawa} \supset \quad & 
\mathbf{Y}_u H_uqu^c + \mathbf{Y}_dH_d^*qd^c+
\mathbf{Y}_e H_d^*e^c\ell +\text{h.c.}\\
\xrightarrow[]{\langle H_{u,d} \rangle = \frac{1}{\sqrt{2}}v_{u,d}}&~
\mathbf{M}_u uu^c + \mathbf{M}_ddd^c+
\mathbf{M}_e e^ce + \text{h.c.}
\end{align}
where $\mathbf{M}_u=\mathbf{Y}_uv_u/\sqrt{2}$ and $\mathbf{M}_{d,e}=\mathbf{Y}_{d,e}v_d/\sqrt{2}$. The SM fermion masses are therefore generated in the usual way in a Type-II 2HDM. Tracing the Yukawa couplings back to their UV origin, we can derive the following relations\footnote{Here we use $\mathbf{Y}_d$ instead of $\mathbf{Y}_e$ for convenience. Within the $SU(5)$ GUT framework, we expect $\mathbf{Y}_d\sim\mathbf{Y}_e$ up to GUT breaking effects. While not precisely identical, they match at the order-of-magnitude level.}
\begin{align}
\mathbf{Y}_H\sim \mathbf{Y}_u=\sqrt{2}\,\mathbf{M}_u/v_u~~,\quad
\mathbf{Y}_{\Phi^*}\sim \mathbf{Y}_d=\sqrt{2}\,\mathbf{M}_d/v_d~,
\end{align}
up to running effects, implying that they are both hierarchical and approximately diagonal $3\times3$ matrices. However, these relations implicitly assume that the masses are generated at tree level directly from the VEVs $v_u$ and $v_d$, which need not be the case. For example, the down-type quark and charged lepton masses can be generated radiatively from $H_u$ \cite{Graham:2009gr}, in which case the relation $\mathbf{Y}_d\propto\mathbf{M}_d$ would no longer hold, and $\mathbf{Y}_{\Phi^*}\sim\mathbf{Y}_d$ could have a significantly different flavor structure.

For the neutrino masses, the relevant terms are instead included in eq.~\eqref{DSS} and eq.~\eqref{PSF}, where $\Delta_0$ generates the RH Majorana neutrino masses at the seesaw scale, while $H_d$ generates the Dirac neutrino masses at the electroweak scale as
\begin{align}\label{NuMass}
-{\cal L}_{\rm Yukawa} \supset ~ 
\mathbf{Y}_\Delta \Delta_0\nu^c\nu^c + \left(\mathbf{Y}_\nu H_d \nu^c\ell+\text{h.c.}\right)
\xrightarrow[~\langle \Delta_0 \rangle = v_R/\sqrt{2}]{~\langle H_d \rangle = v_d/\sqrt{2}}~
\mathbf{M}_R\nu^c\nu^c + \left(\mathbf{M}_\nu \nu^c\nu+\text{h.c.}\right)
\end{align}
where $\mathbf{M}_R=\mathbf{Y}_\Delta v_R/\sqrt{2}$ and $\mathbf{M}_\nu=\mathbf{Y}_\nu v_d/\sqrt{2}$. The tiny neutrino masses can be explained through the Type-I seesaw mechanism with $v_R\sim\Lambda_{\rm Seesaw}\sim 10^{10-14}$ GeV. The resulting light neutrino mass matrix is given by
\begin{align}\label{nuMass}
\mathbf{m}_\nu=-\mathbf{M}_\nu\,{\mathbf{M}_R^{-1}}\,{\mathbf{M}_\nu^T}~.
\end{align}

Here, we obtain another relation, $\mathbf{Y}_{\Phi}\sim \mathbf{Y}_\nu$, since the $H_d\nu^c\ell$ term originates from the $\mathbf{\Phi}\mathbf{S'}\overline{\mathbf{F}}$ term. However, since the Dirac neutrino mass $\mathbf{M}_\nu$ does not directly correspond to the physical neutrino mass, this relation does not provide additional information. The Yukawa coupling $\mathbf{Y}_{\Phi}\sim \mathbf{Y}_\nu$, which is independent of $\mathbf{Y}_u$ and $\mathbf{Y}_d$, can therefore have a completely different flavor structure.

\paragraph{The dark fermion masses}~\\
 
Next, we turn to the dark fermion masses. The VEV of $\phi_0$ not only generates mass terms for the exotic fermions $\psi_c$ and $\psi_w$, but also for all the dark quarks $\chi$ as
\begin{align}
-{\cal L}_{\rm Yukawa} \supset  ~& 
\phi_0^*\left(\mathbf{Y}_a \chi_a\chi^c_a + \mathbf{Y}_c \psi_c\psi_c^c+
\mathbf{Y}_w \psi_w\psi_w^c \right) + \phi_0\left(\mathbf{Y}_s \chi_s\chi_s^c + \mathbf{Y}_f \chi_f\chi_f^c \right)+\text{h.c.}\nonumber\\
\xrightarrow[]{~\langle \phi_0 \rangle = f/\sqrt{2}~ }&~\mathbf{M}_a \chi_a\chi^c_a + \mathbf{M}_c \psi_c\psi_c^c+ \mathbf{M}_w \psi_w\psi_w^c+ \mathbf{M}_s \chi_s\chi_s^c + \mathbf{M}_f \chi_f\chi_f^c +\text{h.c.}
\label{DSMass}\end{align}
where $\mathbf{M}_i=\mathbf{Y}_i f/\sqrt{2}$. Similar to the SM fermions, the dark fermions arising from the extension also come in three generations and feature nontrivial flavor structures.

Tracing these Yukawa couplings back to their UV origin, we obtain two relations
\begin{align}
(1)~\mathbf{Y}_{\Phi^*}:~\mathbf{Y}_a\sim\mathbf{Y}_c\sim\mathbf{Y}_w\sim \mathbf{Y}_{d}=\sqrt{2}\,\mathbf{M}_d/v_d~~,\quad
(2)~\mathbf{Y}_{\Phi}:~\mathbf{Y}_s\sim\mathbf{Y}_f\sim\mathbf{Y}_{\nu}~,
\end{align}
up to symmetry breaking effects. The first relation, if $\mathbf{Y}_{d}=\sqrt{2}\,\mathbf{M}_d/v_d$ holds, has important implications for the dark fermion spectrum, with $\mathbf{M}_a\sim\mathbf{M}_c\sim\mathbf{M}_w\sim (f/v_d)\times\mathbf{M}_d$, which will be discussed in section \ref{sec:Portal}. On the other hand, the second relation does not provide much information beyond $\mathbf{Y}_s\sim\mathbf{Y}_f$. However, even this similarity might not hold if $\phi_0$ has a dynamical origin, as discussed in section \ref{sec:ScalarSB}. For example, if the diagonal breaking is realized by the $\chi_s \psi_f^c$ condensate of $SU(N)'$, the resulting $\phi_0$ can be viewed as a composite bound state of $\chi_s\chi_s^c$, which couples strongly to $\chi_s$ but weakly to $\chi_f$, leading to $\mathbf{Y}_s\gg\mathbf{Y}_f$ and potentially also spoiling the relation $\mathbf{Y}_a\sim\mathbf{Y}_c\sim\mathbf{Y}_w$.

There is one last mass term for the dark fermions, which arises from the Higgs doublet
\begin{align}\label{MixMass}
-{\cal L}_{\rm Yukawa} \supset 
\mathbf{Y}_m H_d^* \chi_f^c\psi_w + \mathbf{\widetilde{Y}}_{m} H_d \chi_f\psi_w^c
\xrightarrow[]{~\langle H_{d} \rangle = v_d/\sqrt{2}~}&~
\mathbf{M}_m \chi_f^c\psi_{w,0} + \mathbf{\widetilde{M}}_m \chi_f\psi_{w,0}^c 
\end{align}
where $\mathbf{M}_{m}(\mathbf{\widetilde{M}})=\mathbf{Y}_m(\mathbf{\widetilde{Y}}_{m})v_d/\sqrt{2}$. It introduces mixing between $\chi_f$ and the neutral component of the doublet $\psi_{w,0}\subset\psi_w$, resulting in a $6\times6$ mass matrix for the $\{\chi_f,\psi_{w,0}\}$ states. If we assume $\mathbf{M}_w\gg\mathbf{M}_m,\mathbf{\widetilde{M}}_m$, then $\mathbf{M}_w$ is approximately unchanged after the mixing, while the mass matrix of the dark quark becomes
\begin{align}\label{Mf}
\mathbf{M}_f=\mathbf{Y}_f \frac{f}{\sqrt{2}}~\to~
\mathbf{M}_f=\mathbf{Y}_f \frac{f}{\sqrt{2}}-\mathbf{\widetilde{M}}_m^\dagger \mathbf{M}_w^{-1} \mathbf{M}_{m}
=\mathbf{Y}_f \frac{f}{\sqrt{2}}-\mathbf{\widetilde{Y}}_m^\dagger \mathbf{Y}_w^{-1} \mathbf{Y}_{m}\frac{v_d^2}{\sqrt{2}f}~,
\end{align}
where the second term can be viewed as a contribution induced by the Higgs VEV $v_d$ after integrating out the heavy $\psi_{w,0}$. Again, tracing the Yukawa couplings back to their UV origin gives the relations $\mathbf{Y}_m \sim \mathbf{Y}_{\Phi^*}\sim \mathbf{Y}_w$ and $\mathbf{\widetilde{Y}}_{m}\sim\mathbf{Y}_{\Phi}\sim \mathbf{Y}_f$. If both relations applied, the eq.~\eqref{Mf} can be simplified to $\mathbf{M}_f\sim(\mathbf{Y}_f-(v_d^2/f^2)\mathbf{Y}_f^\dagger)f/\sqrt{2}$. Although the correction is suppressed by $v_d^2/f^2$ and its contribution to the dark fermion masses is subleading, the interaction term in eq.~\eqref{MixMass} is closely related to the Higgs portal, which will be discussed further in section~\ref{sec:EWportal}.

Now that the complete scalar sector and its interactions have been established, the model is complete. We then move on to discuss the $U(1)$ symmetry structure of the extension in section \ref{sec:U(1)}, the details of the strongly coupled dark sector in section \ref{sec:DS}, and finally the portal interactions between the dark and visible sectors in section~\ref{sec:Portal}.

%\newpage

\subsection{$U(1)$ symmetries and exotic dark baryon number $\widetilde{D}$}\label{sec:U(1)}

One important consequence of a dark sector originating from an extension of the SM is the emergence of new symmetries, especially $U(1)$ symmetries that can play an important role in either dark matter stability or the generation of an asymmetric dark matter number density. In the SM, two accidental symmetries emerge from the bottom-up perspective, namely the well-known \textbf{baryon number $\bm{B}$} and \textbf{lepton number $\bm{L}$}. While both are known to be anomalous, only the combination $\bm{B-L}$ remains anomaly-free and corresponds to a good quantum number. Such a combination becomes apparent when going beyond the SM to grand unified theories, where $B-L$ can arise as a gauge symmetry. In this subsection, we aim to identify analogous $U(1)$ symmetries in our model.

In our model, since all the interaction terms respect the maximal gauge symmetry, the information about these $U(1)$ global symmetries can be traced back to the gauge group $SU(N+5) \times SU(N+1)' \times U(1)''$. In section \ref{sec:ScalarSB}, we already identified four relevant $U(1)$ symmetries in the UV, including
\begin{align}\label{4U(1)}
U(1)_{N},~U(1)_Y\subset SU(N+5)~,\quad
U(1)'_{N}\subset SU(N+1)'~,\quad
U(1)''~.
\end{align} 
Here, $U(1)_Y$ represents the familiar SM hypercharge. Besides, a specific linear combination, $U(1)_D \subset U(1)_{N} \times U(1)'_{N} \times U(1)''$ as defined in eq.\eqref{U(1)D}, also remains unbroken after the symmetry breaking. Given the charge assignments in eq.~\eqref{Dcharge}, under which only the dark fermions are charged, we identify it as the \textbf{dark baryon number $\bm{D}$}, which is an exact $U(1)$ gauge symmetry in our setup.

Next, we try to identify the analogous symmetry to the $B-L$ symmetry. In the standard $SU(5)$ GUT with RH neutrinos, it is related to the $U(1)_X$ symmetry following $B-L=(4Y+X)/5$, under which the fermions $\{\mathbf{10}_F,\overline{\mathbf{5}}_F,\nu^c\}$ carry charges $\{1,-3,5\}$. In our extended framework, we expect a similar conserved $U(1)$ symmetry to emerge, which also encompasses the dark fermions. 

We find a new $U(1)_{X'}$ symmetry with a generator defined by
\begin{align}\label{U(1)X'}
T_{X'}=\frac{1}{2} \left(T_N + T'_N + 2\times T''\right)~,
\end{align} 
which reproduces the correct $B-L$ charges for the SM fermions. We derive the extended version $(B-L)-D'=(4Y+X')/5$ with the fermion charges written in matrix form as
\begin{align}\label{B_L}
(B-L)-D':\,\mathbf{A}=
\begin{pmatrix}
\frac{N+6}{5}  & \frac{3N+13}{30}   &  \frac{N+11}{10} \\
   &   -\frac{1}{3}      &  \frac{1}{3}      \\
   &       &  1    \\
\end{pmatrix}
,~
\overline{\mathbf{F}} =
\begin{pmatrix}
-\frac{N+6}{5}  &  -\frac{N+11}{10}\\ 
-\frac{3N+13}{30}  &  -\frac{1}{3}   \\ 
-\frac{N+11}{10}  &  -1  \\
\end{pmatrix}
,~
\mathbf{S'} =
\begin{pmatrix}
\frac{N+6}{5}      &   \frac{N+11}{10}     \\
   &   1       \\
\end{pmatrix}
\end{align}
One important feature is that the dark fermions also carry charges under this new $U(1)$. We therefore define a new dark baryon number $D'$ to characterize their charges under this symmetry. This gives the correct charge assignment for the SM fermion sector, while the dark fermions acquire somewhat less intuitive charge assignments, which, however, do not affect the conservation of the symmetry.

However, to make the discussion easier and clearer, we consider a linear combination of $(B-L)-D'$ with the unbroken dark baryon number $D$. This allows us to remove the $N$ dependence by defining a new exotic dark baryon number $\widetilde{D}=D'-\frac{N(N+11)}{10}D$ such that the fermion charges under the resulting $U(1)$ symmetry are
\begin{align}\label{B-L-Dfermion}
(B-L)-\widetilde{D}:~\mathbf{A}=
\begin{pmatrix}
-1  & -\frac{2}{3}   &  0 \\
   &   -\frac{1}{3}      &  \frac{1}{3}      \\
   &       &  1    \\
\end{pmatrix}
~,\quad
\overline{\mathbf{F}} =
\begin{pmatrix}
1  &  0\\ 
\frac{2}{3}  &  -\frac{1}{3}   \\ 
0  &  -1  \\
\end{pmatrix}
~,\quad
\mathbf{S'} =
\begin{pmatrix}
-1     &   0     \\
   &   1       \\
\end{pmatrix}~,
\end{align}
where the linear combination is chosen such that the fundamental dark quark carries zero charge under this symmetry\footnote{Another linear combination with $\widetilde{D}=D'-\frac{N(N+6)}{10}D$, where the antisymmetric dark quark $\chi_a$ carries zero charge, was studied in \cite{Chung:2024nnj} due to the Majorana nature of $\chi_a$ is the $Sp(4)_D$ dark QCD theory.}. Notice that this replacement relies on the existence of an unbroken $U(1)_D$, which might not be the case if $U(1)_D$ is broken in a particular model.

We can also check the charges of the relevant scalars under the $(B-L)-\widetilde{D}$ symmetry,
\begin{align}\label{B-L-Dscalar}
(B-L)-\widetilde{D}:~
\Delta_0=-2
,~
\mathbf{\Phi} =
\begin{pmatrix}
0  &  -1\\ 
\frac{1}{3}  &  -\frac{2}{3}   \\ 
1  &  0  \\
\end{pmatrix}
,~
\mathbf{5}_H=
\begin{pmatrix}
-\frac{2}{3} \\ 0
\end{pmatrix}
,~
\overline{\mathbf{10}}_H=
\begin{pmatrix}
\frac{4}{3} &  \frac{2}{3}  \\
    &  0  \\
\end{pmatrix}~.
\end{align}
For the scalars that acquire nonzero VEVs, $\phi_0$ and the Higgs are neutral under $(B-L)-\widetilde{D}$, while $\Delta_0$ carries charge $-2$, thereby breaking the $(B-L)-\widetilde{D}$ symmetry by two units. Apart from this spontaneous breaking, the symmetry is preserved by the Yukawa and gauge interactions, since it originates from a gauge symmetry at the high scale.

The dark fermion charges of the newly defined \textbf{exotic dark baryon number $\bm{\widetilde{D}}$} are
\begin{align}
\widetilde{D}(\chi_f)=\widetilde{D}(\psi_w)=0~,\quad \widetilde{D}(\chi_a)=\widetilde{D}(\chi_s)=1~,\quad \widetilde{D}(\psi_c)=\frac{2}{3}~,
\end{align}
which are clearly distinct from the dark baryon number $D$ we derived previously! If the scalar and gauge bosons that mediate processes violating $L$, $B$, and $\widetilde{D}$ are sufficiently heavy, $B$ and $\widetilde{D}$ can behave as approximately conserved global symmetries, making the proton, as well as the lightest exotic dark baryon with $\widetilde{D}\neq0$, long-lived. In contrast, $L$ is by the Majorana mass term of the neutrinos under our assumption. The complete table of fermion charges is shown in table \ref{tab:U(1)}.

Notice that the conventional dark baryons with $D\neq0$ are already protected by the dark baryon number $D$, which, if it remains an unbroken gauge symmetry at low energies, can render the lightest dark baryon exactly stable. Here, we are instead discuss about the exotic dark baryons with $\widetilde{D}\neq0$ but $D=0$ (thus not protected by $U(1)_D$). These include chimera baryons such as $\chi_a\chi_f^c\chi_f^c$ and $\chi_s\chi_f^c\chi_f^c$. A distinctive feature of these baryons is that they are composed of dark quarks and anti-dark quarks, with their dark color indices contracted by a symmetric tensor, in contrast to conventional baryons, whose dark color indices are contracted by an antisymmetric tensor. In this study, due to their nonzero exotic dark baryon number and unconventional baryon definition, we therefore refer to them as \textbf{exotic dark baryons}. 

\begin{table}
\centering
\begin{tabular} {|c|c|c|c|c|c|c|c|c|c|c|c|c|}\hline
     & $\chi_f$ & $\chi_a$  & $\chi_s$ & $\psi_c$ & $\psi_w$ & $q$ & $u^c$ & $d^c$ & $\ell$  & $e^c$  & $\nu^c$    \\ \hline
${D}$  & $\frac{1}{N}$ & $\frac{2}{N}$ & $\frac{2}{N}$ & $\frac{1}{N}$ & $\frac{1}{N}$ & $0$ & $0$ & $0$ & $0$ & $0$ & $\frac{1}{2}$  \\ \hline
$\widetilde{D}$  & $0$ & $1$ & $1$ & $\frac{2}{3}$ & $0$ & $0$ & $0$ & $0$ & $0$ & $0$ & $0$  \\ \hline
${B}$  & $0$ & $0$ & $0$ & $0$ & $0$ & $\frac{1}{3}$ & $-\frac{1}{3}$ & $-\frac{1}{3}$ & $0$ & $0$ & $0$  \\ \hline
${L}$  & $0$ & $0$ & $0$ & $0$ & $0$ & $0$ & $0$ & $0$ & $1$ & $-1$ & $-1$  \\ \hline
$Y$    & $0$ & $0$ & $0$ & $-\frac{1}{3}$ & $\frac{1}{2}$ & $\frac{1}{6}$ & $-\frac{2}{3}$ & $\frac{1}{3}$ & $-\frac{1}{2}$ & $1$ & $0$ \\ \hline
\end{tabular}
\caption{\label{tab:U(1)}The charges of fermions under the global and gauge $U(1)$ symmetries, $D$, $\widetilde{D}$, $B$, $L$, $Y$.}
\end{table}

In a strongly coupled vector-like dark QCD theory, as considered in this study where we aim to develop a general framework, these exotic dark baryons have a mass origin similar to that of conventional baryons, with their masses set by the dark QCD confinement scale. However, in a strongly coupled \textit{chiral} theory, they can instead appear as massless composite fermions that match the ’t Hooft anomalies \cite{Dimopoulos:1980hn,Bars:1981se}. This leads to distinct properties and can provide a solution to the dark matter-baryon coincidence problem, as discussed in \cite{Chung:2024ezq}. We leave a study of the strongly coupled chiral theory version for future work.

The $(B-L)-\widetilde{D}$ symmetry might also play an important role in the early Universe, particularly in the generation of dark matter. Since the required thermal history depends on the mass and properties of the dark matter candidate, we consider one specific example based on asymmetric dark baryon dark matter to illustrate how this $U(1)$ symmetry can be utilized, as follows.

As we already include the Type-I seesaw mechanism with heavy right-handed neutrinos, the framework can naturally accommodate Leptogenesis \cite{Fukugita:1986hr}. It is well known that this mechanism can generate the required baryon asymmetry. Similarly, in this extended model, the decay of heavy right-handed neutrinos can also be responsible for generating a dark baryon asymmetry. From the Yukawa interactions introduced in section \ref{sec:Yukawa}, we find a decay chain of the form $\nu^c \to \chi_f \phi_\nu^* \to \chi_f\chi_f\chi_{a,s}^c$ that violates both lepton number and exotic dark baryon number. If the decay process violates CP symmetry due to non-vanishing CP phases in the relevant Yukawa couplings and occurs through out-of-equilibrium decays, we can then realize a leptogenesis-like mechanism for the exotic dark baryons. The consequence is that we obtain an asymmetry with more anti-$\chi_{a,s}$ than $\chi_{a,s}$, while there are more $\chi_f$ than anti-$\chi_f$. After dark QCD confinement, these quarks and anti-quarks will form the bound states, in this case, the exotic dark baryons $\chi_{a,s}^c\chi_f\chi_f$. The amount of dark baryon asymmetry generated through this process can be comparable to the lepton asymmetry generated from $\nu^c \to \ell H$ if the Yukawa coupling matrix is similar, which is the case as both interactions originate from the same Yukawa interaction term $\mathbf{\Phi}\mathbf{S'}\overline{\mathbf{F}}$ in eq.~\eqref{PSF}.

Given the diverse interaction terms introduced between the dark sector and the SM, we expect that various dark matter production mechanisms can be realized under this framework, especially those requiring additional symmetries, such as asymmetric dark matter models \cite{Kaplan:2009ag,Davoudiasl:2012uw,Petraki:2013wwa,Zurek:2013wia}. Furthermore, since we have not specified the strong dynamics at this stage, the model can also accommodate weakly coupled dark sectors with the $SU(N)$ gauge group serving an entirely different purpose.

%\newpage

\section{The $SU(N)$ Dark Sector}\label{sec:DS}

In this section, we focus on the dark sector and its low-energy dark hadron spectrum. We start with the dark color gauge group. In this study, we focus on the $SU(N)_D$ dark color group, which originates from the diagonal subgroup of $SU(N)\times SU(N)'$. Other types of dark color groups, such as $Sp(N)_D$ (with an even $N$) \cite{Bennett:2023wjw,Landini:2020daq,Zierler:2021cfa,Kulkarni:2022bvh,Zierler:2022uez,Dengler:2024maq} and $SO(N)_D$ \cite{Hietanen:2012sz,Antipin:2015xia}, can also be obtained with additional symmetry breaking steps. We will not discuss these possibilities in this study, but mention them to emphasize the applicability of our model to diverse dark QCD setups.

For the dark quarks arising from the SM extension, unlike conventional dark QCD models which usually assume only Dirac dark quarks in the fundamental representation, our model contains three types of dark quarks, $\chi_{f,I}$, $\chi_{a,I}$, and $\chi_{s,I}$, transforming as fundamental, antisymmetric, and symmetric representations, respectively, under the $SU(N)_D$ dark color group, each with three generations $I=1,2,3$. In principle, one can also have dark-colored scalars at low energies, such as ${\phi_\nu}^i\subset \mathbf{\Phi}$ and $\tilde{\ell}^c_{0,ijk},~\sigma_{ijk\ell}\subset \mathbf{H}$, which can serve as bosonic dark quarks and make the dark sector even richer. However, we leave these possibilities for future work and focus only on fermionic dark quarks in this analysis.

Next, with the $SU(N)_D$ dark color group and its dark quark content determined, we examine the running of the dark color coupling $g_D$ and the dark color confinement scale $\Lambda_{\rm DC}$. Assuming that the scalar particles and exotic fermions are heavy, only the dark quarks remain at low energies. The one-loop RGE at the scale $\Lambda$ is given by
\begin{align}\label{1loop}
\left.\frac{1}{g_D^3}\frac{d\,g_D}{d\,\text{ln}\,\mu}\right|_\Lambda
=-\frac{1}{16\pi^2}\,b_D ~,~
b_D=\frac{11}{3}N-\frac{4}{3}\left(\frac{1}{2}\,n_{f,\Lambda}+\frac{N-2}{2}\,n_{a,\Lambda}+\frac{N+2}{2}\,n_{s,\Lambda}\right)~,
\end{align}
where $n_{f,\Lambda}$, $n_{a,\Lambda}$, and $n_{s,\Lambda}$ are the numbers of dark quark generations in the corresponding representations below the scale $\Lambda$. In the massless limit, we have $n_{f,\Lambda}=n_{a,\Lambda}=n_{s,\Lambda}=3$, leading to $b_D=-{N}/{3}-2$, which implies that the theory does not confine at low energies. This means that a strongly coupled dark sector containing all the dark quarks is not possible, and at least some of the dark quarks must acquire masses above $\Lambda_{\rm DC}$ to make the coefficient $b_D$ positive.

In our model, we have already shown in eq.~\eqref{DSMass} that all the dark quarks can acquire masses from the VEV $\langle\phi_0\rangle=f/\sqrt{2}$ and may have a hierarchical spectrum similar to the SM fermions. We therefore expect that the theory can eventually enter the confining phase at a sufficiently low energy with some heavy dark quarks decouple.\footnote{For certain numbers of flavors, the theory could reach an infrared fixed point \cite{Sannino:2004qp,Dietrich:2006cm}, featuring unique properties that could be helpful for model building. However, in this study, we pursue a confining theory and therefore require the number of flavors to lie below the conformal window, which is a stronger requirement than merely being asymptotically free.} To remain general, we stay agnostic about the dark quark spectrum and assume the numbers of dark quarks in each representation below the confinement scale $\Lambda_{\rm DC}$ to be $n_f$, $n_a$, and $n_s$, respectively, i.e., we define $n_i\equiv n_{i,\Lambda_{\rm DC}}$ for the following discussion.

With a positive $b_D$, the dark color coupling increases toward lower energies, causing the theory to become strongly coupled and the dark quarks to form condensates at the confinement scale $\Lambda_{\rm DC}$. Below the confinement scale, the dark gluons and dark quarks are no longer appropriate degrees of freedom. Instead, the system should be described in terms of hadronic degrees of freedom, including dark glueballs, dark mesons, and dark baryons.

The properties of dark hadrons depend on the nature of the dark color group and how the dark quarks transform under it, which in general requires a dedicated lattice study for each scenario. In this study, we do not aim for a comprehensive study of the $SU(N)$ dark sector, but instead highlight some distinct features of the dark sector in our model, especially the two types of dark quarks, $\chi_a$ and $\chi_s$, with unconventional representations, and provide some existing lattice references. For this purpose, we focus on dark hadrons formed from the dark quarks and refer the discussion of dark glueballs to other dedicated studies \cite{Soni:2016gzf,Forestell:2016qhc,Carenza:2022pjd,Carenza:2023eua,Batz:2023zef,McKeen:2024trt}. We also ignore the special case with $n_f=n_a=n_s=0$, corresponding to the heavy-quark limit with $m_\chi\gg\Lambda_{\rm DC}$, which has been studied in \cite{Mitridate:2017oky,Assi:2023cfo,DallaValleGarcia:2026aws,DallaValleGarcia:2026mis}.

Since the representation theory with $N<5$ is rather special and less general, we will go through $N=2,3,4$ one by one and then consider $N\geq5$ in the following subsections. For each case, we start with the dark quark properties, showing the properties of the different representations. From that, we discuss the pattern of chiral symmetry breaking and the resulting dark pions, which are the lightest dark hadrons composed of dark quarks. Finally, some properties of dark baryons are presented.

\subsection{$N=2$}

Starting from the $SU(2)_D$ dark color theory, the dark quarks of different representations have the following properties
\begin{align}
\chi_f=\mathbf{2}~{\rm (pseudo-real)}~,\quad \chi_a=\mathbf{1}~{\rm (singlet)}~,\quad
\chi_s=\mathbf{3}~{\rm (real)}~,
\end{align}
In this case, we can simply ignore the antisymmetric representation $\chi_a$, which is a singlet under $SU(2)_D$, and the symmetric representation $\chi_s$ under $SU(2)_D$ is equivalent to the adjoint representation. This multiple-representation fermion setup was originally proposed in the context of walking technicolor \cite{Ryttov:2008xe}, and subsequently explored in lattice studies \cite{Bergner:2020mwl,Bergner:2021ivi}. In this section, we briefly summarize its relevant properties and refer interested readers to these original works for a more comprehensive discussion.

With numbers of dark quarks $n_f$ and $n_s$, the dark sector features a global symmetry $SU(2n_f)\times SU(2n_s)$, where the subscripts denote two different representations. Once the dark color becomes strongly coupled and the dark quark condensates, $\chi_f\chi_f^c$ and $\chi_s\chi_s^c$, form, the system becomes confining, and we need to describe it in terms of hadronic degrees of freedom, namely dark mesons and dark baryons.

Starting with the mesons, multiple types of mesons are possible, depending on the dark quark constituents, either $\chi_f$ or $\chi_s$, and their quantum numbers, including pseudoscalar, scalar, vector, tensor, axial-vector, and axial-tensor mesons. For phenomenological interest, we only focus on the lightest states, the pseudoscalar mesons, usually called dark pions, while other states typically have masses of order the dark color confinement scale $\Lambda_{\rm DC}$. Since pseudoscalar mesons are known as the Nambu-Goldstone bosons of global symmetry breaking, we therefore keep track of the symmetry breaking pattern.

In this case, the $\langle\chi_f\chi_f^c\rangle$ and $\langle\chi_s\chi_s^c\rangle$ condensates lead to the following breakings
\begin{align}
SU(2n_f)\to Sp(2n_f)~,\quad SU(2n_s)\to SO(2n_s)
\end{align}
respectively. We can then count a total of $(2n_f^2-n_f-1)$ $\pi_f$'s from the coset $SU(2n_f)/Sp(2n_f)$ and $(2n_s^2+n_s-1)$ $\pi_s$'s from the coset $SU(2n_s)/SO(2n_s)$, where we label the dark pions with subscripts denoting their quark constituents.

Regarding the dark baryons, depending on the representations of their constituents, there are three types of baryons: fundamental-only $\chi_f\chi_f$ ($D=1$), symmetric-only $\chi_s\chi_s$ ($D=2$), and chimera baryons $\chi_s\chi_f\chi_f$ ($D=2$) composed of mixed representations. In a $SU(N)$ theory, we expect higher-dimensional representations to form a condensate at a higher scale due to a smaller required coupling. Therefore, a greater dynamical (constituent) mass is expected for $\chi_s$ than for $\chi_f$. And the lightest baryon should be the spin-1 bound state of $\chi_{f}$ as studied in \cite{Francis:2018xjd}. If we further take the hierarchical fermion spectrum into account and focus on the one-flavor theory with only the lightest dark quark $\chi_{f,1}$, we find the lightest baryon to be a $\rho$ vector diquark/meson transforming as a triplet under the unbroken $Sp(2)f(=SU(2)f)$ flavor symmetry
\begin{align}\label{rho}
\rho=
\begin{pmatrix}
\rho^+    \\   \rho^0    \\   \rho^- \\
\end{pmatrix}\sim
\begin{pmatrix}
~\chi_{f,1}\chi_{f,1}~\\\chi_{f,1}^c\chi_{f,1}\\ \chi_{f,1}^c\chi_{f,1}^c \\
\end{pmatrix}~\text{with}~
D=
\begin{pmatrix}
+1    \\   0    \\   -1 \\
\end{pmatrix}~.
\end{align} 
where we also show their $U(1)_D$ charge which is a gauged subgroup of $Sp(2)_f$. Notice that, due to the pseudo-real nature of $\chi_f$ under $SU(2)_D$, baryons (diquarks) with $D=1$ and mesons with $D=0$ share identical color structures and live within the same multiplets under the extended flavor symmetry.

\subsection{$N=3$}

Next, $N=3$ is the same gauge group as the SM QCD, while the dark quarks include
\begin{align}
\chi_f=\mathbf{3}~{\rm (complex)}~,\quad \chi_a=\mathbf{\bar{3}}~{\rm (complex)}~,\quad
\chi_s=\mathbf{6}~{\rm (complex)}~.
\end{align}
In this case, $\chi_a$ is the anti-fundamental representation and has the same behavior as $\chi_f$. The main difference from QCD comes from the additional sextet representation, which has been studied on the lattice, with a special focus on $n_s=2$ \cite{Shamir:2008pb,DeGrand:2008kx,Fodor:2009ar,DeGrand:2009hu} due to its potential near-conformal behavior. In this study, we focus on the confining phase and the resulting hadron properties.

With numbers $n_f$, $n_a$, and $n_s$, the system features an enhanced global symmetry of $SU(n_f+n_a)_L\times SU(n_f+n_a)_R \times SU(n_s)_L \times SU(n_s)_R$ symmetry, which is, however, broken by the $U(1)_D$ gauge symmetry. Once the dark color becomes strongly coupled and the dark quark condensate forms, the global symmetry undergoes spontaneous symmetry breaking to the subgroup $SU(n_f+n_a)_V \times SU(n_s)_V$, giving rise to $(n_f+n_a)^2-1$ dark pions from the triplet fermions and $n_s^2-1$ dark pions from the sextet fermions. Among them, however, the $2n_fn_a$ dark pions of mixed representations $\chi_f\chi_a$ receive masses of $\mathcal{O}(g_D\Lambda_{\rm DC}/4\pi)$ from the $U(1)_D$ gauge boson loop, reflecting its explicit breaking effect.

For the baryon spectrum, we now have baryons of one, two, and even three species of fermion. For baryons composed of the same species, we have $B_f=\chi_f\chi_f\chi_f$, $B_a=\chi_a\chi_a\chi_a$, and $B_s=\chi_s\chi_s\chi_s$. Again, we expect $B_s$ to be heavier, while for $B_f$ and $B_a$, though both by triplet fermions, they have different charges under $U(1)_D$, $D(B_f)=1$ and $D(B_a)=2$, so if the dark quark masses are the same, $B_a$ should be heavier due to the $U(1)_D$ contribution. Furthermore, if we also consider baryons of mixed fermion representations, one can get exotic dark baryons $\chi_a^c\chi_f\chi_f$ and $\chi_s^c\chi_f\chi_f$ featuring dark baryon number $D=0$ but $\widetilde{D}=1$, which could be the lightest dark baryon. There is even a chimera baryon composed of three distinct fermions, $\chi_s\chi_a\chi_f^c$, which is a unique state only in the $SU(3)$ theory.

\subsection{$N=4$}

For the $N=4$ case, the dark quarks transform as
\begin{align}
\chi_f=\mathbf{4}~{\rm (complex)}~,\quad 
\chi_a=\mathbf{6}~{\rm (real)}~,\quad
\chi_s=\mathbf{10}~{\rm (complex)}~.
\end{align}
A distinctive feature of this case is that the antisymmetric representation is real. This scenario is probably one of the most extensively studied gauge theories with mixed representations due to its application as a UV completion for composite Higgs models \cite{Ferretti:2013kya,Ferretti:2014qta}, motivating numerous related lattice investigations \cite{DeGrand:2015lna,Ayyar:2017qdf,Ayyar:2018zuk,Cossu:2019hse}.

Now the fundamental and symmetric representations, which are complex, feature a common chiral symmetry $SU(n_i)_L\times SU(n_i)_R$ which is broken down to $SU(n_i)_V$, with $n_i^2-1$ dark pions. Meanwhile, the antisymmetric representation, which is real, features an enhanced global symmetry $SU(2n_a)$, which is broken down to $SO(2n_a)$, introducing $(2n_a^2+n_a-1)$ dark pions.

For the baryon spectrum, we have baryons composed of the same dark quarks, including $B_f=\chi_f\chi_f\chi_f\chi_f$, $B_a=\chi_a\chi_a$, and $B_s=\chi_s\chi_s\chi_s\chi_s$. Notice that the diquark $B_a$ here, due to the real nature of $\chi_a$ under $SU(4)_D$, shares identical color structures with the $\chi_a^c\chi_a$ meson and lives in the same multiplets under the extended flavor symmetry, similar to the $\chi_f$ hadrons we discuss in the $N=2$ case. Beyond the conventional baryons, there is again chimera baryons with two distinct fermions, such as $\chi_a\chi_f\chi_f$, which lives in the same $SO(2n_a)$ multiplets as the exotic dark baryons $\chi_a^c\chi_f\chi_f$.

\subsection{$N\geq5$}

For large $N\geq5$, the condition becomes rather trivial as
\begin{align}
\chi_f=\ydiagram{1}~{\rm (complex)}~,\quad 
\chi_a=\ydiagram{1,1}~{\rm (complex)}~,\quad
\chi_s=\ydiagram{2}~{\rm (complex)}~.
\end{align}
All the representations are complex and independent. After forming the condensate, they undergo the same chiral symmetry breaking pattern $SU(n_i)_L\times SU(n_i)_R\to SU(n_i)_V$ and introduce $n_i^2-1$ dark pions for each species.

For baryons, conventional baryons composed of $N$ dark quarks are still present. On top of these, we again have chimera baryons. We emphasize again the exotic dark baryons $\chi_a^c\chi_f\chi_f$ and $\chi_s^c\chi_f\chi_f$, featuring dark baryon number $D=0$ but carrying $\widetilde{D}=1$. This kind of dark baryon is especially mentioned in strongly coupled chiral gauge theories \cite{Dimopoulos:1980hn,Bars:1981se}, for example in $SU(5)$ with $\mathbf{10}+\mathbf{\bar{5}}$, as a light composite fermion.

%\newpage

\section{Portal with the SM}\label{sec:Portal}

\begin{figure}[tbp]
\centering
\includegraphics[width=1.0\textwidth]{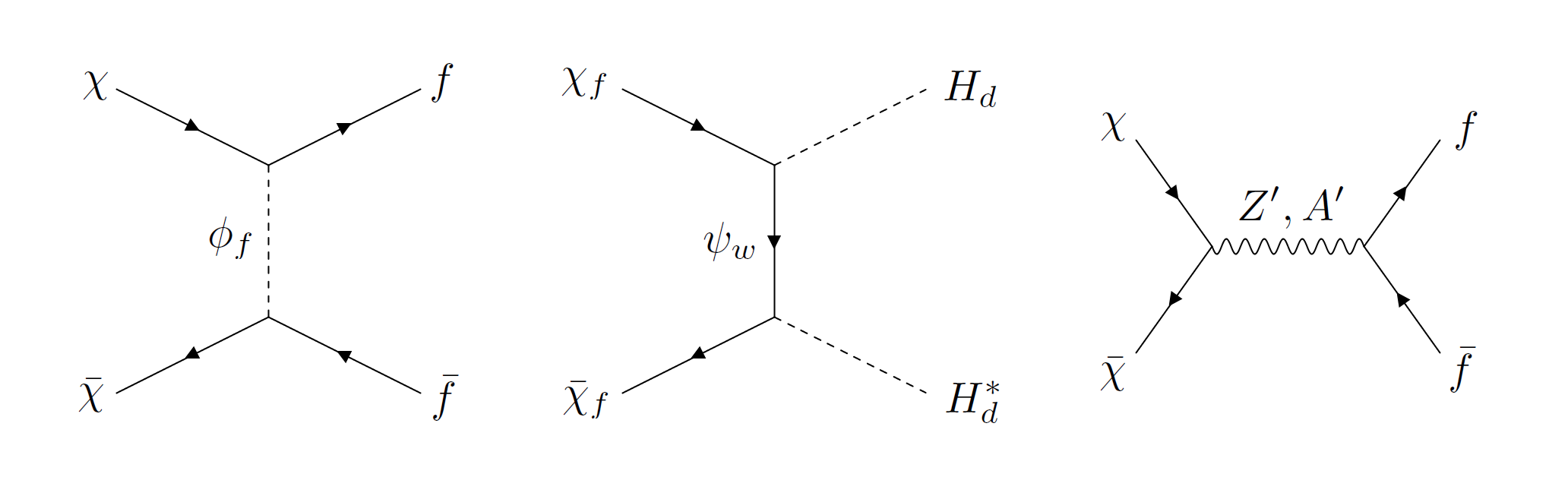}
\caption{Interactions between the dark quarks $\chi$ and the SM particles through the exchange of heavy scalars (left), heavy fermions (middle), and neutral gauge bosons (right). \label{fig:Portal}}
\end{figure}

Having settled the $SU(N)$ dark sector, we then turn to how it interacts with the SM particles. From the Yukawa and gauge interactions discussed in section \ref{sec:Model}, we can identify various interaction terms that connect the dark sector to the visible sector. In this section, we systematically categorize and present these interactions according to their origin and portal type, focusing on the states expected to be relatively light and therefore most relevant for low energy phenomenology. 

We start with the three portal interactions arise at tree level as shown in figure \ref{fig:Portal}, including the quark/lepton portal by heavy scalars in section \ref{sec:Scalar}, the Higgs/$Z$ portal by heavy fermions in section \ref{sec:EWportal}, and the $Z'/A'$ portals by new gauge bosons in section \ref{sec:Gauge}. We then compare these interactions in section \ref{sec:Compare}, especially the coefficients of various dimension-six operators. Other portals arising at one loop are discussed in section \ref{sec:Other}. Direct searches for selected mediators are discussed in section \ref{sec:direct_search}.

In the analysis, we simply write down the operators arising from these interactions and discuss some of their distinct properties, especially some unique prediction from our framework, such as the Yukawa couplings and gauge charges. To preserve the broad applicability of our framework, we leave a comprehensive phenomenological study for future work once a specific dark sector is chosen, but we mention some dedicated studies that correspond to specific scenarios. We do, however, highlight the key interactions involving dark pions, the lightest dark hadronic states under our consideration\footnote{In this work, we focus on dark sectors with at least one dark quark present below the confinement scale $\Lambda_{\rm DC}$; for the heavy-quark limit, see instead \cite{Mitridate:2017oky,Assi:2023cfo,DallaValleGarcia:2026aws,DallaValleGarcia:2026mis}. Even with $n_\chi=1$, for which the dark pion shift symmetry is anomalous, lattice studies show that the dark pion (also known as the $\eta$ meson in the one-flavor theory) is the lightest hadron \cite{Bennett:2023rsl,Francis:2018xjd}.}, which behave similarly to conventional fermionic axion-like particles (ALPs) and naturally govern the core low energy phenomenology of the dark sector.

To be consistent with most phenomenological studies, we will switch to Dirac spinor notation for all new equations, while retaining the Weyl spinor notation for the interaction terms already presented.

\subsection{Heavy scalar mediator}\label{sec:Scalar}

Start with the Yukawa interactions between the dark quarks and the SM fermions. As we have already listed the complete Yukawa interaction terms in section \ref{sec:Yukawa}, we only show the relevant terms in this section. Before moving to the actual interaction terms in our model, we first discuss the general impact of a $t$-channel heavy scalar mediator. A general Yukawa interaction term between one type of SM chiral fermion $f\in\{q,u^c,d^c,\ell,e^c\}$ and one type of dark chiral fermion $\chi\in\{\chi_f,\chi_f^c,\chi_a,\chi_a^c,\chi_s,\chi_s^c\}$ through a scalar $\phi_f$ can be written as
\begin{align}
\mathcal{L}_{\rm Yukawa}=\mathbf{Y}_{IJ}\phi_ff_I\chi_J+\text{h.c.}~.
\end{align}
Since both sectors come in three generations\footnote{The fact that the dark sector and the SM fermions share the same number of flavors, $n_\chi=n_f=3$, actually restricts the properties of the dark sector. For example, in \cite{Carmona:2024tkg}, the stable dark pion arises from an enhanced symmetry for $n_\chi>3$, which cannot be realized in our model.}, the Yukawa coupling has a nontrivial flavor structure described by a $3\times3$ matrix labeled by the flavor indices $I,J=1,2,3$, which is important for phenomenology, just as flavor is in the SM.

Such an interaction can lead to direct detection of dark matter through the effective four-fermion operators
\begin{align}
{\cal L}_{\rm four-fermion} = 
\frac{\mathbf{Y}^{\dagger}_{I'I}\mathbf{Y}_{J'J}}{2M_\phi^2}
\left(\bar{\chi}_{I'}\gamma^\mu \chi_{J'}\right)
\left(\bar{f}_I\gamma_\mu f_J\right)~,
\end{align}
with the cross section controlled by $\mathbf{Y}^4/M_\phi^4$. For a dark QCD theory, at low energies after dark color confinement, dark pions arise as bound states of $\bar{\chi}\chi$, which can also interact with the SM through the same operator, given by
\begin{align}
{\cal L}_{\pi_D} 
=\mathbf{Y}^{\dagger}_{I'I}\mathbf{Y}_{J'J}\frac{f_D}{2M_\phi^2}\partial^\mu\pi_{D,I'J'} \left(\bar{f}_I\gamma_\mu\gamma_5 f_J\right)~,
\end{align}
where $f_D$ is the decay constant of the dark pion $\pi_D$. The interaction term is the same as that of a fermionic axion-like particle (ALP) \cite{Dolan:2014ska,Dobrich:2018jyi,Bauer:2020jbp,Bauer:2021mvw,DallaValleGarcia:2023xhh,Ovchynnikov:2025gpx}, but with the interaction scale $f_a\propto M_\phi^2/f_D$ and depends on the flavor-nonuniversal Yukawa couplings.

In the subsections, we divide our discussion into the quark portal ($f \in \{q, u^c, d^c\}$) and the lepton portal ($f \in \{\ell, e^c\}$). As we will see, only interactions between specific dark quarks and SM fermions arise under our minimal scalar-sector setup. We further discuss the implications of our UV model for the two portals.

%\newpage

\subsubsection{Quark Portal}\label{sec:Quark}

Starting from quark-philic interactions \cite{Bai:2013iqa}, the quark portal dark QCD sector, mediated by a heavy colored scalar, has been extensively explored in the literature \cite{Renner:2018fhh,Mies:2020mzw,Bensalem:2021qtj,Carmona:2021seb,Carmona:2022jid,Carmona:2024tkg}. These models are of particular interest due to their phenomenological implications at the Large Hadron Collider (LHC), notably the generation of emerging jet signatures \cite{Schwaller:2015gea,Archer-Smith:2021ntx,Carrasco:2023loy}. Theoretically, such a colored scalar is also motivated to reside at the TeV scale in order to break the coupled infrared fixed points of QCD and dark QCD \cite{Bai:2013xga,Newstead:2014jva,Ritter:2022opo,Ritter:2024sqv}.

In our model, colored scalars with similar interactions also arises naturally from the extension of the SM. The three relevant interaction terms are given by
\begin{align}
\mathbf{Y}_{df}\phi_dd^c\chi_f,\quad \mathbf{Y}_{ua}\phi_uu^c\chi_a,\quad \mathbf{Y}_{qa}\phi_qq\chi_a~.
\end{align}
We can see that only specific combinations of fermions arise. In particular, the right-handed down-type quark $d^c$ couples only to the left-handed fundamental dark quark $\chi_f$, while the left-handed antisymmetric dark quark $\chi_a$ can couple to both $u^c$ and $q$. The right-handed dark quarks $\chi_f^c$ and $\chi_a^c$, as well as the symmetric dark quarks $\chi_s,\chi_s^c$, are not involved. This is a consequence of the Yukawa interactions arising from the minimal scalar sector. One can introduce all possible combinations by extending the scalar sector beyond the minimal setup. However, introducing additional scalars would only lead to redundant interactions governed by arbitrary couplings. In this work, we therefore follow the minimal setup and retain only the minimal scalar sector, which also provides a more predictive framework.

One of the main predictions of our model from the extension of the SM is the relation between Yukawa couplings. While we have discussed potential effects that could alter this picture in section \ref{sec:Yukawa}, barring such scenarios, our framework yields two relations based on the UV origin of the Yukawa couplings as
\begin{align}
(1)~\mathbf{Y}_{\Phi}:~\mathbf{Y}_{df}\sim\mathbf{Y}_\nu,\quad (2)~\mathbf{Y}_{H}:~\mathbf{Y}_{ua}\sim\mathbf{Y}_{qa}\sim\mathbf{Y}_u=\sqrt{2}\,\mathbf{M}_u/v_u~.
\end{align}
The first relation is probably not so useful, as we do not know the structure of $\mathbf{Y}_\nu$. The second one, however, predicts a hierarchical and approximately diagonal Yukawa coupling matrix between the dark quarks $\chi_a$ and SM quarks $u^c,\,q$. This implies strong interactions with the heavy quarks while suppressed interactions with light quarks, making heavy flavor physics a more sensitive and promising channel to test this kind of portal.

Another challenge for the quark-portal scenario and the colored scalar mediator is that, if we adopt the scalar spectrum assumption (2) in eq.~\eqref{Mphi2} and eq.~\eqref{Mphi2*}, which assumes that the mechanism making the color triplet Higgs $T_u$ and $T_d$ heavier than the weak doublets $H_u$ and $H_d$, i.e., the solution to the doublet-triplet splitting problem, also applies to the colored scalars discussed here, they can then be as heavy as the GUT scale, making the quark portal less relevant. Under this assumption, one should instead focus on the colorless scalars and the lepton portal.

%\newpage

\subsubsection{Lepton Portal}\label{sec:Lepton}

The colorless scalar state can lead to Yukawa couplings between dark quarks and SM leptons, giving rise to lepton portal dark matter \cite{Bai:2014osa}. Compared to the colored scalar case, a lepton portal dark QCD sector has been less extensively studied \cite{Zhang:2021orr,Lu:2025cty}, partly due to the lack of a clear UV motivation. However, in our model, since the scalars originate from the extension of the SM Higgs sector, the symmetry, such as the $SU(N+1)$ symmetry relating $\phi_\ell$ and $H_d$ within $\mathbf{\Phi}$ in eq.~\eqref{Phi}, can allow the colorless scalar to remain as light as the electroweak scale. This makes the lepton portal a more natural and phenomenologically viable possibility, and it may even be accessible in direct searches by LHC, as discussed in section \ref{sec:direct_search}.

From the interaction term in section \ref{sec:Yukawa}, we find two relevant Yukawa interactions,
\begin{align}
\mathbf{Y}_{\ell f}\phi_\ell\ell\chi_f,\quad \mathbf{Y}_{ea}\phi_ee^c\chi_a
\end{align}
Again, we find a specific combination, the left-handed lepton $\ell$ coupling to the left-handed fundamental dark quark $\chi_f$ and the right-handed charged lepton $e^c$ coupling to the left-handed antisymmetric dark quark $\chi_a$, while the other dark quarks are not involved. The interaction structure closely parallels that of the quark portal. This naturally follows from the $SU(5)$ assignments in eq.~\eqref{SU(5)F}, where $\ell$ and $d^c$ arise from the same fiveplet, and $e^c$ resides in a decuplet alongside $u^c$ and $q$.

The relation between the Yukawa couplings can again be derived from their UV origin as
\begin{align}
(1)~\mathbf{Y}_{\Phi}:~\mathbf{Y}_{\ell f}\sim\mathbf{Y}_\nu~,\quad 
(2)~\mathbf{Y}_{H}:~\mathbf{Y}_{ea}\sim\mathbf{Y}_u=\sqrt{2}\,\mathbf{M}_u/v_u~.
\end{align}
Similarly, we find the first relation less useful, as we do not know much about $\mathbf{Y}_\nu$. The second relation, however, predicts a hierarchical and approximately diagonal Yukawa coupling matrix between the dark quarks $\chi_a$ and the right-handed charged lepton $e^c$.

There are a few more interesting Yukawa interactions regarding leptons
\begin{align}
\mathbf{Y}_{\Delta}\left(\Delta_1\nu^c\chi_f+\Delta_2\nu^c\chi_s\right),\quad 
\mathbf{Y}_{\nu f}\phi_\nu\nu^c\chi^c_f
\end{align}
which couple the right-handed neutrino $\nu^c$ to different dark quarks. In our model, we already include the Type-I seesaw, with right-handed neutrino masses at $10^{10-14}$ GeV, and therefore we do not expect $\nu^c$ to be relevant at low energies. However, these interactions can still play a role in the early Universe, such as in dark matter genesis through leptogenesis, as briefly illustrated at the end of section \ref{sec:U(1)}.

While we emphasize the scalar spectrum following assumption (2) in eq.~\eqref{Mphi2} and eq.~\eqref{Mphi2*} to motivate the lepton portal, there is also a possibility that all these dark-colored scalar masses are dominated by the $\phi_0$'s VEV at the diagonal scale, as in assumption (1) in eq.~\eqref{Mphi1} and eq.~\eqref{Mphi1*}. In that case, the two portals can be comparable in strength, leading to rather rich dark sector phenomenology.

%\newpage

\subsection{Heavy doublet fermion: Higgs and $Z$ portal}\label{sec:EWportal}

The portal introduced by the Yukawa interaction terms is not limited to fermion portals, since we also have a scalar field in the SM, the Higgs field, which couples with the light dark quarks $\chi_f$ and the heavy doublet fermions $\psi_w$. The relevant interaction term was already mentioned in eq.~\eqref{MixMass} in the context of mixing between $\chi_f$ and $\psi_{w,0}$. Here, we emphasize its role as the origin of the Higgs and $Z$ portals. The relevant interaction terms include
\begin{align}\label{LHP}
-{\cal L}_{\rm Yukawa} \supset\mathbf{Y}_mH_d^*\psi_w\chi_f^c+\mathbf{\widetilde{Y}}_{m}H_d\chi_f\psi_w^c+\mathbf{Y}_w \phi_0^* \psi_w\psi_w^c+\text{h.c.}~.
\end{align}
Once $\phi_0$ acquires a VEV with $\langle \phi_0 \rangle = f/\sqrt{2}$ at the diagonal breaking scale, the exotic doublet fermions become massive with $\mathbf{M}_w=\mathbf{Y}_w f/\sqrt{2}$. We then obtain the tree-level diagram shown in figure~\ref{fig:Portal} (middle).

If the doublet fermions $\psi_w$ are much heavier than the electroweak scale and the dark QCD scale, we can then integrate them out, resulting in the following effective interaction, where we switch to Dirac spinor notation as
\begin{align}\label{HP}
\mathcal{L}_\text{Higgs portal}= \frac{\sqrt{2}}{f}\bar{\chi}_{f,L}\,\mathbf{\widetilde{Y}}_m^\dagger \mathbf{Y}_w^{-1} \mathbf{Y}_{m} \,\chi_{f,R} |H_d|^2+\text{h.c.}~.
\end{align}
This is a dimension-five Higgs portal operator for the fundamental dark quarks $\chi_f$ and the down-type Higgs doublet $H_d$, featuring a characteristic scale $f$.

We can further check the relation between the Yukawa coupling matrices based on their UV origin, which should satisfy
\begin{align}\label{YYM}
(1)~\mathbf{Y}_{\Phi^*}:~\mathbf{Y}_m\sim \mathbf{Y}_{w}\sim \mathbf{Y}_{d}=\sqrt{2}\,\mathbf{M}_d/v_d~,\quad 
(2)~\mathbf{Y}_{\Phi}:~\mathbf{\widetilde{Y}}_{m}\sim \mathbf{Y}_\nu,
\end{align}
The combination $\mathbf{Y}_w^{-1} \mathbf{Y}_{m}$ in eq.~\eqref{HP} then cancels, becoming $\sim \mathbb{I}_{3\times3}$ . The resulting interaction is nevertheless controlled by the rather unknown coupling $\mathbf{\widetilde{Y}}_{m}\sim \mathbf{Y}_\nu$.

This setup is also common in dark QCD models, especially those motivated by the hierarchy problem \cite{Craig:2015pha,Graham:2015cka,Cheng:2018gvu,Cheng:2019yai,Cheng:2021kjg}. We emphasize the study \cite{Cheng:2021kjg}, from which we borrow the same notation, while we have additional guidance on the Yukawa matrices from eq.~\eqref{YYM}. Compared with the previous quark and lepton portals, the Higgs portal arises from a dimension-five operator, which implies less suppression. Moreover, if $\mathbf{M}_w=\mathbf{Y}_w f/\sqrt{2}$ follows the hierarchical structure of $\mathbf{M}_{d}$, the interaction mediated by the lightest $\psi_{w,1}$ can be quite enhanced, rendering it an even more promising portal.

The operator also has a broad range of phenomenological implications. First, it can give rise to direct detection signals mediated by the Higgs. Second, if the dark QCD sector has dark hadrons lighter than the Higgs mass, one can get Higgs invisible decay with a strong bound of Br$(h\to {\rm invisibles})<10.7\%$ at the 95\% CL \cite{ATLAS:2023tkt}. Lastly, CP-even dark pions can also mix with the Higgs boson directly through this operator.

From the same interaction terms in eq.~\eqref{LHP}, we also get dimension-six $Z$ portal operators
\begin{align}
\mathcal{L}_\text{$Z$ portal}= -\frac{g_Zv_d^2}{2f^2}Z_\mu
\left(\bar{\chi}_{f,L}\mathbf{\widetilde{Y}}_m^\dagger \mathbf{Y}_w^{-2} \mathbf{\widetilde{Y}}_{m} \gamma^\mu\chi_{f,L}-
\bar{\chi}_{f,R}\mathbf{Y}_m^\dagger \mathbf{Y}_w^{-2} \mathbf{Y}_{m}\gamma^\mu\chi_{f,R}
\right)
\end{align}
which, on the other hand, can induce mixing between the $Z$ boson and the CP-odd dark pions. Although relatively suppressed compared with the dimension-five Higgs portal, the $Z$ portal can play an important role in specific cases \cite{Cheng:2019yai}.

Note that, given the Yukawa interaction terms we introduce, this portal only arises for the $H_d\subset \mathbf{\Phi}$, the Higgs doublet that couples to the down-type quarks and charged leptons. Therefore, the exact strength of these two portals depends strongly on the potential of two Higgs doublets, which determines the parameters $\alpha$ and $\beta$ and is not specified in this study. However, it is worth mentioning that there is a scenario in which $H_d$ only acquires a tiny VEV, $v_d\simeq 0$, with the down-type quarks and charged leptons instead acquiring masses from $H_u$ radiatively \cite{Graham:2009gr}, resulting in $H_{\rm SM} \simeq H_u$ with a very large $\tan\beta$, known as the uplifted Higgs region of parameter space \cite{Dobrescu:2010mk,Altmannshofer:2010zt}. In this parameter space, both the Higgs and $Z$ portals discussed above might become highly suppressed. Moreover, relation (1) in eq.~\eqref{YYM} will no longer hold.

\subsection{Gauge bosons from the extended gauge group}\label{sec:Gauge}

So far, we have discussed the portals arising from the heavy scalars and their Yukawa interactions in section \ref{sec:Yukawa}. There are also additional gauge interactions arising from the extension of the SM to the maximal gauge group discussed in section \ref{sec:gauge}. Among them, many can mediate interactions between the SM and the dark sector.

However, in this work, we follow the symmetry breaking chain and corresponding scales discussed in section \ref{sec:ScalarSB} and focus only on the gauge interactions that can remain relevant at low energies. Therefore, we do not discuss the gauge bosons that acquire masses at the first three scales, $\Lambda_{\rm DGUT}$, $\Lambda_{\rm GUT}$, and $\Lambda_{\rm Seesaw}$, which we expect to be well above $10^{10}$ GeV or even higher. Instead, we focus on the gauge group in eq.~\eqref{leftover}. The breaking at $\Lambda_{\rm Diagonal}$ gives rise to multiple massive gauge bosons whose masses can be much lighter than those associated with the first three scales. Ignoring the gauge bosons that mediate interactions only within the dark sector, we are left with a heavy $Z'$ neutral gauge boson that couples to both sectors, as discussed in section \ref{sec:Z'}.

Besides, the unbroken $U(1)_D$ gauge group introduces a massless dark photon. Although at tree level it interacts only within the dark sector, kinetic mixing between the dark photon and the SM photon can be generated at loop level \cite{Holdom:1985ag}, providing another popular and promising portal covered in section \ref{sec:A'}.

%\newpage

\subsubsection{Heavy $Z'$ portal}\label{sec:Z'}

The heavy $Z'$ boson at the diagonal breaking scale $\Lambda_{\rm Diagonal}$ arises from the breaking of $U(1)_N\times U(1)''_N\to U(1)_D$, while the properties of $U(1)''_N$ should be traced back to the breaking of $U(1)'_N\times U(1)''\to U(1)''_N$ at the seesaw scale. To derive the details, we define the couplings of $U(1)_N$, $U(1)'_N$, $U(1)''$ as $g_N$, $g'_N$, $g''$, respectively, adopting the generators defined in section \ref{sec:ScalarSB}, the corresponding covariant derivative is
\begin{align}
D_\mu=\partial_\mu-ig_NZ_{N,\mu} T_N-ig'_NZ'_{N,\mu} T'_N-ig''Z''_\mu T''
\end{align}
With the coupling and the generator of each $U(1)$, we can then derive the properties of the $Z'$ boson following the $U(1)$ symmetry breaking chain. The detailed calculation is presented in Appendix \ref{App:U(1)SB}, and here we only summarize the relevant results.

First, the mass of the $Z'$ boson is given by
\begin{align}\label{Z'mass}
M_{Z'}=5 g_{Z'}f ~, \quad {\rm where}~~g_{Z'}=\sqrt{g_N^2+\frac{(N+1)^2{g'_N}^2{g''}^2}{(2Ng'_N)^2+((N+5)g'')^2}}
\end{align}
The large prefactor in this expression is due to the normalization we choose. The couplings of the $Z'$ boson are dictated by the covariant derivative
\begin{align}\label{Z'coupling}
D_\mu=\partial_\mu-ig_{Z'}Z'_\mu\left(T_N-N(N+5)\,{\rm sin}^2\theta~ T_D\right)~,
\end{align}
where $T_N$ is given in eq.~\eqref{GammaVEV} and $T_D$ is given in eq.~\eqref{U(1)D}. The mixing angle $\theta$ is given by 
\begin{align}\label{theta}
{\rm tan}\theta=\frac{N+1}{g_N}
\frac{g'_Ng''}{\sqrt{(2Ng'_N)^2+((N+5)g'')^2}}~.
\end{align}

From eq.~\eqref{Z'coupling}, we see that the coupling between the $Z'$ boson and the dark fermions receives contributions from both the $T_N$ and $T_D$ terms, and therefore depends on the mixing angle $\theta$ and, consequently, on the exact values of the three couplings appearing in eq.~\eqref{theta}. On the other hand, since the SM fermions do not carry $U(1)_D$ charge, the coupling between the $Z'$ boson and the SM fermions is determined solely by the $T_N$ term. The relative strength is then determined by the SM fermion charges under $U(1)_N$, which are given by
\begin{align}
Q'(u^c):Q'(q):Q'(e^c):Q'(d^c):Q'(\ell)=(-2):(-2):(-2):1:1~.
\end{align}

One attractive consequence is that, assuming the $Z'$ is the dominant portal between the dark sector and the SM fermions, the coupling strength to the different SM fermions should follow this ratio. For a dark pion coupling to the SM fermions through the $Z'$ boson, the effective operator is then given by (in Dirac-spinor notation)
\begin{align}
{\cal L}_{\pi_D}
=\frac{\partial^\mu\pi_D}{f_a} 
\left(4\,\bar{u}\gamma_\mu\gamma_5 u
+\bar{d}\gamma_\mu\gamma_5 d
+\bar{e}\gamma_\mu\gamma_5 e
-\bar{\nu}\gamma_\mu\gamma_5 \nu\right)~,
\end{align}
where the scale $f_a\propto f^2/f_D$. The interaction is flavor-universal but species-dependent, with the relative couplings uniquely determined by the charge ratio given above, and can therefore result in unusual decay properties and branching ratios.

%\newpage

\subsubsection{Kinetic mixing portal}\label{sec:A'}

The breaking at the diagonal breaking scale leaves an unbroken $U(1)_D$ symmetry, which can have a mass much lighter than the $Z'$ boson or even remain massless, and thus can have a significant phenomenological impact. At tree level, the dark photon $A'$ only interacts with dark fermions. However, kinetic mixing with the SM photon can arise at one loop through exotic fermions carrying electromagnetic charge \cite{Holdom:1985ag}.

With a UV model, a clear origin of kinetic mixing can be traced back to the fiveplet $\psi=(\psi_c,\psi_w)$, which transforms as $(\mathbf{5},\frac{1}{N})$ under $SU(5)\times U(1)_D$. The kinetic mixing then originates from the mass splitting between $\psi_c$ and $\psi_w$, which can arise from differences in their Yukawa coupling running. The expected kinetic mixing parameter $\epsilon$ is then given by
\begin{align}
{\cal L}=-\frac{\epsilon}{2} F'_{\mu\nu}F^{\mu\nu}~,\quad
\epsilon\sim\frac{1}{16\pi^2}\frac{eg_D}{N}\frac{\Delta M_\psi}{M_\psi}~,~{\rm where}\quad
\frac{\Delta M_\psi}{M_\psi}\equiv\frac{\frac{1}{3}{\rm Tr}\,\mathbf{M}_c-\frac{1}{2}{\rm Tr}\,\mathbf{M}_w}{\frac{1}{3}{\rm Tr}\,\mathbf{M}_c+\frac{1}{2}{\rm Tr}\,\mathbf{M}_w}
\end{align}
and the coupling $g_D$ is given in eq.~\eqref{gD}. The kinetic mixing should be zero in the UV, where the fiveplet masses are degenerate. After including the effects of $SU(5)$ breaking, however, the mass splitting arises and leads to a nonzero $\epsilon$ effectively at the two-loop level. We therefore expect a typical size of $\epsilon\sim 10^{-4}-10^{-6}$ \cite{Arkani-Hamed:2008kxc,Gherghetta:2019coi}.

Conventional dark baryons in our dark QCD sector carry nonzero $U(1)_D$ charge and can therefore couple to the SM through the dark photon. While the dark matter can also consist of exotic dark baryons with $D=0$, such as $A\bar{F}\bar{F}$ \cite{Chung:2024ezq}, whose stability is guaranteed by nonzero ${\widetilde{D}}$ instead. Since the constituent dark quarks still carry $U(1)_D$ charges, the exotic dark baryon can still interact with the dark photon through dipole interactions \cite{Kuwahara:2026hpr}, similar to the case of the dark neutron \cite{Figueroa:2026bmi}.

Even without the presence pf $U(1)_D$ dark photon at low energies, the vector mesons of the strongly coupled dark sector can also have kinetic mixing with the SM photon \cite{Alonso-Alvarez:2023rjq}. The required ingredients to realize this portal also exist in our model.

%\newpage

\subsection{Comparison between different portals}\label{sec:Compare}

\begin{table}[]
\centering
\begin{tabular}{|c|c|c|c|c|c|}
\hline
Portal  & interaction & effective operator & dim. & mediator  & (coupling)$^2$  \\ \hline
\multirow{3}{*}{Quark} 
& $\mathbf{Y}_{df}\phi_dd^c\chi_f$  & $(\bar{\chi}_{f,L}\gamma^\mu \chi_{f,L})(\bar{d}_R\gamma_\mu d_R)$  & $6$   & $\phi_d$  & $\mathbf{Y}_{df}^2\sim\mathbf{Y}_{\nu}^2$  \\ \cline{2-6}
& $\mathbf{Y}_{ua}\phi_uu^c\chi_a$  & $(\bar{\chi}_{a,L}\gamma^\mu \chi_{a,L})(\bar{u}_R\gamma_\mu u_R)$  & $6$   & $\phi_u$  & $\mathbf{Y}_{ua}^2\sim\mathbf{Y}_{\nu}^2$  \\ \cline{2-6}
& $\mathbf{Y}_{qa}\phi_qq\chi_a$    & $(\bar{\chi}_{a,L}\gamma^\mu \chi_{a,L})(\bar{q}_L\gamma_\mu q_L)$      & $6$   & $\phi_q$  & $\mathbf{Y}_{qa}^2\sim\mathbf{Y}_{u}^2$  \\ \hline
\multirow{2}{*}{Lepton} 
& $\mathbf{Y}_{\ell f}\phi_\ell\ell\chi_f$&$(\bar{\chi}_{f,L}\gamma^\mu \chi_{f,L})(\bar{\ell}_L\gamma_\mu \ell_L)$& $6$ & $\phi_\ell$  & $\mathbf{Y}_{\ell f}^2\sim\mathbf{Y}_{\nu}^2$  \\ \cline{2-6}
& $\mathbf{Y}_{ea}\phi_ee^c\chi_a$  & $(\bar{\chi}_{a,L}\gamma^\mu \chi_{a,L})(\bar{e}_R\gamma_\mu e_R)$  & $6$   & $\phi_e$  & $\mathbf{Y}_{ea}^2\sim\mathbf{Y}_{u}^2$  \\ \hline
Higgs & $\mathbf{Y}_mH_d^*\psi_w\chi_f^c$   & $(\bar{\chi}_{f}\chi_{f})|H_d|^2$  & $5$   & $\psi_w$  & $\mathbf{\widetilde{Y}}_m^\dagger\mathbf{Y}_{m}\sim\mathbf{Y}_\nu^\dagger\mathbf{Y}_{d}$  \\ \cline{3-6}
$Z$   & $\mathbf{\widetilde{Y}}_{m}H_d\chi_f\psi_w^c$  & $(\bar{\chi}_{f,L(R)}\gamma^\mu \chi_{f,L(R)})g_Zv_d^2Z_\mu$  & $6$   & $\psi_w$  & $\mathbf{\widetilde{Y}}_m^2(\mathbf{Y}_{m}^2)\sim\mathbf{Y}_{\nu}^2(\mathbf{Y}_{d}^2)$  \\ \hline
$Z'$   & $g_{Z'}Z'_\mu J_{\chi,f}^\mu$  & $(\bar{\chi}\gamma^\mu \chi)(\bar{f}\gamma_\mu f)$  & $6$   & $Z'$  & $g_{Z'}^2Q'_\chi Q'_f$  \\ \hline
$A'$   & $-\frac{\epsilon}{2} F'_{\mu\nu}F^{\mu\nu}$  & $eA'_\mu(\bar{f}\gamma^\mu f)$  & $4$   & $\psi$ (loop)  & $\epsilon\sim\frac{1}{16\pi^2}\frac{eg_D}{N}\frac{\Delta M_\psi}{M_\psi}$  \\ \hline
\end{tabular}
\caption{The summary of different portals, including their effective operators and properties.\label{tab:portal}}
\end{table}

In the previous subsections, we have discussed all possible tree-level portals arising within our framework. In this subsection, we evaluate the relative importance of each portal, which depends on the dimension of the resulting effective operators, the properties of the mediators, and the associated couplings, as summarized in table~\ref{tab:portal}.
Although an exact comparison is highly model-dependent and relies on the specific choice of the dark sector\footnote{Besides, the $Z'$ and $A'$ portals couple to all dark quark species and therefore remain viable regardless of the specific dark quark composition of the dark sector. For example, a dark sector with only light fundamental dark quarks $\chi_f$ can only have additional $d_R$ and $\ell_L$ portals, while light antisymmetric dark quarks $\chi_a$ can have $u_R$, $q_L$, and $e_R$ portals. However, if there is only light symmetric dark quarks $\chi_s$, then only the $Z'$ and $A'$ portals are allowed, restricting its phenomenological implications.}, we provide general insights into their relative strengths guided by the model.

Before proceeding to the comparison, we introduce a few additional assumptions. Although this sacrifices some generality, it simplifies the subsequent discussion. First, as we are interested in the impact on dark sector phenomenology, which is typically dominated by the lightest dark quarks, we therefore focus on the first generation fermions with $I=1$. Similarly, for the SM sector, although heavy flavor physics is known to play an important role in dark sector searches, most direct detection and beam dump experiments are based on first generation SM fermions. We therefore focus on the $(1,1)$ entry of the Yukawa coupling matrix, $\mathbf{Y}_{11}=y_{11}$. To make the comparison possible, we further assume that all Yukawa couplings share the same hierarchical flavor structure as the up-type quark Yukawa coupling $\mathbf{Y}_u$. We thus take $y_{11}\sim m_u/v_{\rm EW}\sim 10^{-5}$ in the following discussion to reflect the Yukawa suppression. This assumption can of course be relaxed, leading to different phenomenological scenarios, but we adopt it throughout this subsection for simplicity and leave a more general discussion of its potential implications to a dedicated study.

We now turn our attention to table~\ref{tab:portal}. As we know, the dimension of the operator dictates the power-law suppression by the new physics scale, thereby governing the low energy phenomenological impact of the corresponding portal. The kinetic mixing portal, which arises at dimension four, clearly stands out from the other portals. Moreover, the dark photon couples to all dark quarks with the coupling $g_D$ given in eq.~\eqref{gD} and the charges specified in eq.~\eqref{Dcharge}, free from any Yukawa suppression. In our framework, kinetic mixing naturally arises at one loop through the electrically charged fermions $\psi=(\psi_c,\psi_w)$, with the mass difference between $\psi_c$ and $\psi_w$ determining the size of $\epsilon$. In our setup, the mass difference $\Delta M_\psi$ arises at one loop from the different contributions of $SU(3)_C$ and $SU(2)_W$, resulting in an effective two-loop suppression, with a typical size of $\epsilon\sim 10^{-4}-10^{-6}$.

The dark photon itself, as a light degree of freedom of the dark sector, can be directly probed in experiments \cite{Fabbrichesi:2020wbt,Caputo:2021eaa,Antel:2023hkf,Caputo:2026pdw}. However, depending on the dark sector spectrum, it might decay predominantly into either visible or invisible final states, leading to very different signatures. Moreover, the lightness of the dark photon is not guaranteed, and the breaking of $U(1)_D$ can occur at scales as high as the Planck scale beyond the minimal model, rendering it phenomenologically irrelevant. For example, if one wants to realize the generation of an asymmetry in conventional dark baryons in this model, $U(1)_D$ must be broken before dark baryogenesis so that the corresponding dark baryon number violation can occur, as required by the Sakharov conditions \cite{Sakharov:1967dj}.

Next, we consider the Higgs portal arising at dimension five suppressed with only a single power of the heavy mass scale, $\mathbf{M}_w^{-1}$, as derived in eq.~\eqref{HP}. This portal is often assumed to be dominant. In our model, however, the coupling $\mathbf{\widetilde{Y}}_m^\dagger\mathbf{Y}_{m}\sim\mathbf{Y}_\nu^\dagger\mathbf{Y}_{d}$ introduces a double Yukawa suppression. Part of this suppression is compensated by the Yukawa coupling that determines the masses of the mediator $\psi_w$. To be more specific, the first generation $\psi_w$ has a mass $m_{w,1}\sim y_{11}f\sim 10^{-5} f$ and could therefore be much lighter than the diagonal breaking scale $\Lambda_{\rm Diagonal}$, as is also the case for $\psi_{c,1}$. This feature avoids the double Yukawa suppression that appears in the heavy scalar case, giving the Higgs portal for the first generation dark quarks as
\begin{align}
{\cal L}_\text{Higgs portal}=C_{\rm HP}(\bar{\chi}_{f,1}\chi_{f,1})|H|^2\quad{\rm with}\quad
C_{\rm HP}\sim\frac{y_{11}^2}{m_{w,1}}\frac{1}{{\rm tan}^2\beta}\sim\frac{10^{-5}}{f}\frac{1}{{\rm tan}^2\beta}
\end{align}
Here we see another source of uncertainty, namely the ${\rm tan}\beta$ factor. As we already mentioned, restricting ourselves to the minimal scalar sector gives only the coupling between $\chi_f$ and $H_d$. If we consider the naive 2HDM with $v_u\sim v_d$, the estimation is reasonable. However, some models predict the uplifted Higgs region of parameter space \cite{Dobrescu:2010mk,Altmannshofer:2010zt} with an unusually large ${\rm tan}\beta$, in which case the Higgs portal is further suppressed.

The rest of the portals introduce dimension-six operators connecting the dark quarks with SM fermions. Although they are typically more suppressed than the kinetic mixing portal and the Higgs portal, if both of the unfortunate conditions mentioned above are realized, dimension-six operators could become the leading contribution, and it is therefore important to compare their relative strengths. Since they have the same dimension, we only need to analyze their couplings and mediator masses, which determine the final Wilson coefficients. Again, we will focus on the operators involving first generation dark quarks and SM fermions.

Starting with the coupling strength, it is obvious that all the portal terms originating from Yukawa interactions suffer from the double Yukawa suppression $\mathbf{Y}_{11}^2=y_{11}^2\sim 10^{-10}$ as shown in the last column of table \ref{tab:portal}. The only exception is the $Z'$ coupling which we expect to be $g_{Z'}^2Q'_\chi Q'_f\sim 10^{-2}$. Note that the large charge is a consequence of our conventional normalization, and we expect the $g_{Z'}$ itself to be relatively small. Instead, the physical combination $g_{Z'}Q'$ should have a typical strength comparable to that of $U(1)_{\rm EM}$, for which we take $g_{Z'}Q'\sim 0.1$.

The next factor controlling the size of the Wilson coefficients is the mediator mass. Starting with the mass we know, the $Z'$ boson mass is derive in Appendix \ref{App:U(1)SB} and shown in eq. \eqref{Z'mass}. The large prefactor is again due to the conventional normalization condition. For simplicity, we just write $M_{Z'}\sim g_{Z'}Q'f$, where $f$ is the scale of $\Lambda_{\rm Diagonal}$. Combining the coupling with the mediator mass, we get the Wilson coefficient of the $Z'$ portal as $1/f^2$, which is a typical behavior for a gauge boson mediator.

On the other hand, the $Z$ portal is determined by the mass of the heavy fermion $\psi_w$, which also originates from $\langle \phi_0 \rangle = f$, but with a hierarchical flavor structure. We therefore expect the lightest exotic fermion $\psi_{w,1}$ to acquire a mass $\sim y_{11}f \sim 10^{-5}f$, enhancing the strength of the $Z$ portal operator. We already mentioned this effect in the Higgs portal discussion. For the $Z$ portal, which is suppressed by two powers of the heavy mass scale $\mathbf{M}_w^{-2}$ due to its dimension-six nature, the enhancement is squared, and the Yukawa suppression is canceled out, leaving the Wilson coefficient of order $1/f^2$, the same as that of the $Z'$ portal. However, as $v_d$ appears explicitly in the operator, we might still have an additional $1/{\rm tan}^2\beta$ suppression.

Lastly, we have several operators generated by integrating out the scalar $\phi$'s with different charges. While these operators suffer from double Yukawa suppression, they can still play a role if the corresponding scalars are sufficiently light compared with the scale $f$. In this study, however, we do not address the details of the scalar potential, as it is generically unconstrained. This remains true even for scalars arising from the same multiplet. For example, we have $\{\phi_d,\phi_\ell,\phi_0,H_d,T_d\}\subset\mathbf{\Phi}$ within the same multiplet, where $\phi_0$ acquires a VEV $f$ that sets the diagonal breaking scale $\Lambda_{\rm Diagonal}$, whereas $H_d$ should have a VEV around or below the electroweak scale $\Lambda_{\rm EW}$. Moreover, the absence of proton decay signals requires $T_d$ to be as heavy as the GUT scale $\Lambda_{\rm GUT}$, leading to the well-known doublet-triplet splitting problem. It is therefore difficult to determine the mass scales of the scalars. Here, we simply follow the two assumptions mentioned in section \ref{sec:ScalarSB}.

The first assumption in eq.~\eqref{Mphi1} and eq.~\eqref{Mphi1*} simply assumes that they all acquire masses from $\phi_0$ at the diagonal breaking scale, with $M_{\phi_i}\sim f$, especially $\phi_d$ and $\phi_\ell$, which are in the same multiplet as $\phi_0$. It basically implies that the scalar potential among the dark-colored scalars is still governed by the $SU(N+5)$ symmetry. The other scalars, $\phi_u$, $\phi_q$, and $\phi_e$, might also acquire masses from $\phi_0$ through the potential $\mathbf{H}^\dagger\mathbf{\Phi}^\dagger\mathbf{\Phi}\mathbf{H}$. Under this assumption, however, the operators suffer from double Yukawa suppression, resulting in Wilson coefficients $\sim y_{11}^2/f^2\sim 10^{-10}/f^2$, which are much smaller than those of the $Z'$ and $Z$ portals.

The other assumption, as in eq.~\eqref{Mphi2} and eq.~\eqref{Mphi2*}, follows the guidance of doublet-triplet splitting between $T_d$ and $H_d$ and assumes that the same mechanism also applies to the scalar $\phi$'s, leading to the color scalars $\phi_d$, $\phi_u$, and $\phi_q$ being as heavy as $\Lambda_{\rm GUT}$, while $\phi_\ell$ and $\phi_e$ can be around the electroweak scale $\Lambda_{\rm EW}$. Under this assumption, the quark portal becomes irrelevant, while the effective operator for the lepton portal among the first-generation fermions has a Wilson coefficient $\sim y_{11}^2/v_{\rm EW}^2\sim 10^{-10}/v_{\rm EW}^2\sim 1/(10^7\,{\rm GeV})^2$. Therefore, if the diagonal breaking scale $f>v_{\rm EW}/y_{11}\sim 10^7\,{\rm GeV}$, the lepton portal can become the dominant portal.

Finally, we summarize the results in table~\ref{tab:dimsix} and express the Wilson coefficient in term of the suppression scale to reflect the importance of each operators. We emphasize that this comparison is intended primarily to provide general insights, such as the impact of Yukawa suppression and the role played by the scale $f$. Since some of the simplifying assumptions adopted here may not hold in all specific dark sector realizations, a comprehensive exploration of the various portals remains highly motivated and strongly encouraged.

\begin{table}[]
\centering
\begin{tabular}{|c|c|c|c|c|}
\hline
Portal   & effective operator  & mediator (mass)  & coupling & (coefficient)$^{-\frac{1}{2}}$  \\ \hline
\multirow{3}{*}{Quark} 
& $(\bar{\chi}_{f,L}\gamma^\mu \chi_{f,L})(\bar{d}_R\gamma_\mu d_R)$     & $\phi_d(\sim f; \sim \Lambda_{\rm GUT})$  & $y_{11}$  &  $M_{\phi_d}/y_{11}$\\ \cline{2-5}
& $(\bar{\chi}_{a,L}\gamma^\mu \chi_{a,L})(\bar{u}_R\gamma_\mu u_R)$     & $\phi_u(\sim f; \sim \Lambda_{\rm GUT})$  & $y_{11}$  &  $M_{\phi_u}/y_{11}$\\ \cline{2-5}
& $(\bar{\chi}_{a,L}\gamma^\mu \chi_{a,L})(\bar{q}_L\gamma_\mu q_L)$         & $\phi_q(\sim f; \sim \Lambda_{\rm GUT})$  & $y_{11}$  &  $M_{\phi_q}/y_{11}$\\ \hline
\multirow{2}{*}{Lepton} 
& $(\bar{\chi}_{f,L}\gamma^\mu \chi_{f,L})(\bar{\ell}_L\gamma_\mu \ell_L)$&  $\phi_\ell(\sim f; \sim \Lambda_{\rm EW})$  & $y_{11}$  &  $M_{\phi_\ell}/y_{11}$\\ \cline{2-5}
&  $(\bar{\chi}_{a,L}\gamma^\mu \chi_{a,L})(\bar{e}_R\gamma_\mu e_R)$   & $\phi_e(\sim f; \sim \Lambda_{\rm EW})$    & $y_{11}$  &  $M_{\phi_e}/y_{11}$\\ \hline
$Z$    & $(\bar{\chi}_{f,L(R)}\gamma^\mu \chi_{f,L(R)})g_Zv_d^2Z_\mu$  & $\psi_{w,1} (\sim y_{11} f)$    & $y_{11}$  &  $f$\\ \hline
$Z'$   & $(\bar{\chi}\gamma^\mu \chi)(\bar{f}\gamma_\mu f)$    & $Z'(\sim g_{Z'}Q'f)$               & $g_{Z'}Q'$  &  $f$ \\ \hline
\end{tabular}
\caption{Comparison of dimension-six operators coupling dark quarks to SM fermions (first generation). The table lists the mediators along with their masses and couplings. The corresponding Wilson coefficients, expressed in terms of the effective suppression scales, are presented. \label{tab:dimsix}}
\end{table}

%\newpage

\subsection{Other portals}\label{sec:Other}

In this subsection, we briefly discuss several other common portals. While they are as prevalent in the literature as the ones discussed above, in our framework they are induced by the primary interactions either at higher dimensions or at the loop level, and are thus further suppressed. Although they are expected to yield only subleading effects, they can still become phenomenologically relevant in certain specialized scenarios.

\subsubsection{Dipole Portal}\label{sec:Dipole}

The interaction between dark quarks and SM fermions discussed above can also induce an interaction between dark quarks and the SM photon at one loop, leading to a dipole portal \cite{Pospelov:2000bq,Sigurdson:2004zp,Kopp:2009et,Hambye:2021xvd}. Using the heavy scalar $\phi$'s as an example, after integrating out the heavy particles in the loop, we obtain the portal
\begin{align}
{\cal L}_{\rm Dipole} =
d_M \bar{\chi}\sigma^{\mu\nu}\chi F_{\mu\nu}~,\quad
\text{with}\quad
d_M =-\frac{1}{64\pi^2}eQ_f|\mathbf{Y}|^2\frac{m_\chi}{M_\phi^2}
\end{align}
which essentially arises from a one-loop dimension-six operator suppressed by $M_\phi^2$. Here, however, the resulting portal can avoid  the double Yukawa suppression by having the SM third generation fermions in the loop. The operator is even more relevant for lepton portal dark matter, or which the leading direct detection constraint is typically dominated by DM-nucleus scattering through the photon rather than direct DM-electron scattering due to the electron wave-function suppression \cite{Bai:2014osa}.

The dipole operator, after dark quark condensation, can lead to kinetic mixing between the vector meson and the SM photon as
\begin{align}
\langle \omega^\mu| d_M \bar{\chi}\sigma^{\mu\nu}\chi F_{\mu\nu}|0\rangle ~\to~ 
\epsilon_\omega \omega^{\mu\nu}F_{\mu\nu} \quad,~
\text{with}\quad
\epsilon_\omega \sim d_M \Lambda_{\rm DC}
\end{align}
where $\omega^{\mu\nu}$ is the field strength of the meson $\omega^{\mu}$. Therefore, even without the dark photon from $U(1)_D$ gauge symmetry, we still have vector bosons arising as $\bar{\chi}\chi$ mesons from the strongly coupled dark sector, with a tiny kinetic mixing set by the loop factor, Yukawa suppression, and $m\chi\Lambda_{\rm DC}/M_\phi^2$.

\subsubsection{Dark pion/Axion portals}\label{sec:axion}

Although there is no elementary axion-like particle (ALP) in our theory, dark pions arise as the Nambu-Goldstone bosons of the broken global symmetry and behave similarly to ALPs. It is therefore worth discussing the conventional axion portals.

Since dark pions arise as bound states of dark-quark anti-dark-quark pairs, the conventional dimension-five operators for axion portals actually arise from operators of higher dimension as
\begin{align}
c_5\,\partial_\mu \pi_D (\bar{f}\gamma^\mu\gamma^5 f)~\to~ c_6\,(\bar{\chi}\gamma_\mu\gamma_5 \chi)(\bar{f}\gamma^\mu\gamma^5 f) ~,\quad
c'_5\,\frac{\alpha_s}{4\pi}\pi_D G\tilde{G} ~\to~ c'_7\,\frac{\alpha_s}{4\pi}(\bar{\chi}\gamma^5\chi)G\tilde{G}
\end{align}

The first term is precisely the operator we have already introduced in the previous subsection. The four-fermion operator, after dark quark condensation, gives rise to the dark pion-SM fermion operator with $c_5=c_6\times f_D$, where $f_D$ is the dark pion decay constant. It is worth mentioning that, if the scale $f_D$ is as large as the mediator mass, the additional power of the suppression scale will be compensated, rendering the operator effectively dimension-five. Using the $Z'$ portal with $c_6=1/f^2$ as an example, if we have $f=f_D$, we simply obtain the dark pion operator with $c_5=1/f_D$. This scenario is possible if the diagonal breaking is induced by the dark quark condensate.

For the second term, the situation is somewhat different, as the dimension-seven operator arises only at one loop involving a charged fermion and a scalar. In our framework, one can use the interaction in eq.~\eqref{PAF}, which involves the dark quark $\chi_a$, charged fermion $\psi_c$, and scalar $\phi_d$, to construct the desired loop diagram coupling $\bar{\chi}\chi$ to $G\tilde{G}$. Integrating out the heavy fields, we obtain
\begin{align}
\mathcal{L}_\text{axion portal}= \frac{1}{M_{\phi_d}^2}
(\bar{\chi}_a\mathbf{Y}_{\Phi^*}^\dagger \mathbf{M}_c^{-1} \mathbf{Y}_{\Phi^*}\gamma^5\chi_a)\frac{\alpha_s}{4\pi}G\tilde{G}
~\xrightarrow{~\Lambda_{\rm dQCD}~}~
\frac{y^2f_D^2}{M_{\phi_d}^2m_c}\frac{\alpha_s}{4\pi}\pi_a G\tilde{G}~.
\end{align}
The operator is highly suppressed. However, if the dark quark condensate is again responsible for generating the masses of $\psi_c$ and $\phi_d$, such that $M_{\phi_d}\sim m_c\sim f_D$, the coefficient $c'_5$ is determined by $1/f_D$ and it can therefore be as important as other operators.

Here, we show that, in a specific scenario where the relevant scales are determined by the dark QCD confinement scale, the dimension-six and dimension-seven operators can be as important as the dimension-five operators. This, however, implies a rather different model setup that requires dynamical symmetry breaking to play an important role in the symmetry breaking chain, and we leave such a distinct setup for future study.

%\newpage

\subsection{Direct searches of heavy mediators}\label{sec:direct_search}

Our goal in this study is to pursue a list of potential portal terms and their corresponding effective operators after integrating out the heavy mediators. However, we notice that some mediators might be light enough to be searched for directly at the LHC, potentially resulting in even stronger constraints. We emphasize four specific cases, including the scalars $\phi_\ell$ and $\phi_e$, which can be at the electroweak scale if the doublet-triplet splitting assumption applies, and exotic fermions $\psi_{w,1}$ and $\psi_{c,1}$, whose masses might be low due to the hierarchical Yukawa couplings, with $m_{w,1},m_{c,1}\sim y_{11}f\sim 10^{-5}f$. All these exotic particles carry SM charge and can be pair-produced through Drell-Yan processes. Once produced, they can decay through various channels featuring dark quarks and SM particles, leading to interesting signatures.

Start from the two scalars, $\phi_\ell$ and $\phi_e$, that couple to leptons and dark quarks. After being pair produced, each of them decays to a lepton and a dark quark. The dark-quark and anti-dark-quark pair in the final state give rise to two dark jets, which characterize the signal. Assuming the dark pions are sufficiently long-lived to escape the detector, we obtain a signature with opposite-sign dilepton plus missing transverse energy, which is similar to the signature of sleptons in supersymmetric models. With a hierarchical Yukawa coupling, we expect the branching ratio to the $\tau$ final state to be dominant, so the searches for the two scalars $\phi_\ell$ and $\phi_e$ can be matched to $\tilde{\tau}_L$ and $\tilde{\tau}_R$, respectively. Assuming the coupling is large enough to ensure prompt decay, we can therefore apply the latest result by ATLAS using 139 fb$^{-1}$ of Run 2 data~\cite{ATLAS:2024fub}, which gives $M_{\phi\ell}>425$ GeV and $M_{\phi_e}>350$ GeV. It was shown that, for a dark sector with dark matter below $8$ GeV, these collider searches at the LHC can provide the leading constraint compared with xenon-based direct detection experiments \cite{PandaX:2024pjr}.

Next, we turn to exotic fermions. The $\psi_{w,1}$ carrying EW charges has already been studied in \cite{Cheng:2021kjg}. After being pair produced, each of them decays to an electroweak boson and a dark quark, leaving a signature with $WZ/Wh$ plus missing transverse energy, which is similar to the signatures of electroweakinos in supersymmetric models. The current bound comes from ATLAS using 139 fb$^{-1}$ of Run 2 data \cite{ATLAS:2021yqv}, which gives $m_{w,1}\gtrsim 1$ TeV. This bound, following the mass relation $m_{w,1}=y_{11}f\sim 10^{-5}f$ we estimated, then gives $f>10^8$ GeV, implying a rather high diagonal breaking scale.

For $\psi_{c,1}$, which is expected to have a mass comparable to $\psi_{w,1}$, the situation is rather involved because it does not directly couple to a dark quark plus a SM particle. If all the colored scalar fields are heavy under assumption (2) of the scalar spectrum, the most likely decay chain for $\psi_{c,1}$ would be $\psi_{c,1}\to q\phi_\ell \to q\ell \chi_f(ue\chi_f/d\nu\chi_f)$. It therefore depends on the mass hierarchy between $\psi_{c,1}$ and $\phi_\ell$, as well as the mass splitting among the neutral and charged components of $\phi_\ell$. Focusing on the case with a light neutral component of $\phi_\ell$, the final state $d\nu\chi_f$ features a down-type quark with missing transverse energy, similar to down-type squark searches at the LHC. Considering a hierarchical Yukawa coupling matrix with $y_{13}>y_{11}$, the dominant final states then feature $b$-jets with less background, and the constraint can be as strong as $m_{c,1}>1.3$ TeV \cite{ATLAS:2021yij}, which again prefers a value of $f$ as high as $10^8$ GeV following the relation $m_{c,1}=y_{11}f\sim 10^{-5}f$.

In this subsection, we only give a few brief ideas about how the model can have an impact on LHC searches without considering the details or other possible conditions. For example, if the dark pions are not sufficiently long-lived and decay inside the detector, the phenomenology becomes similar to the emerging jets scenario \cite{Schwaller:2015gea,Archer-Smith:2021ntx,Carrasco:2023loy}, but now the signal can arise from electroweak production and be accompanied by leptons or electroweak bosons in the final states. We encourage dedicated studies, especially taking the details of dark sector into account, which might lead to more fruitful results.

%\newpage

\section{Conclusions}\label{sec:Conclusion}

%~\\ - The Idea ~\\

In this work, we have constructed a strongly coupled $SU(N)$ dark sector by extending the Standard Model. This approach is primarily inspired by the fact that SM singlets naturally arise from extensions of the SM fermion sector within the $SU(5)$ GUT framework. The dark sector introduced in this manner inherits the chiral and flavor features of the SM fermions. Specifically, the dark fermions transform under three distinct representations, fundamental, antisymmetric, and symmetric, each possessing exactly three generations. This framework provides a compelling theoretical justification for a dark QCD sector with mixed representations beyond conventional constructions, and also naturally accommodate the rich dark flavor structures widely conjectured in the literature.

%~\\ - The Model ~\\

With the fermion content determined by this extension, we construct a concrete model based on the maximal gauge group $SU(N+5) \times SU(N+1)' \times U(1)''$, where the dark color group $SU(N)_D$ arises as the diagonal subgroup of $SU(N+5) \times SU(N+1)'$. Two intriguing $U(1)$ symmetries emerge from this structure. The first is the dark baryon number $D$, which not only ensures the stability of dark baryons but can also remain unbroken, introducing a dark photon. The second one is the $(B-L)-\widetilde{D}$ symmetry, the extended version of the standard $B-L$ symmetry, which now incorporates the new dark fermions, which carry an exotic dark baryon number $\widetilde{D}$ distinct from $D$.

In the scalar sector, however, we adhere to the principle of minimality by introducing only four necessary scalar multiplets. This minimal set is sufficient to realize all required symmetry breakings from $SU(N+5) \times SU(N+1)' \times U(1)''$ to $G_{\rm SM} \times SU(N)_D \times U(1)_D$, and generate masses for all fermions. Correspondingly, this framework allows for four gauge-invariant Yukawa interaction terms. These terms not only govern mass generation but also introduce portal interactions that are relevant to dark sector phenomenology.

%~\\ - Portal with the SM ~\\

A primary goal of this work is to systematically analyze the diverse portals arising from the extended scalar and gauge sectors. Alongside the conventional quark portal, Higgs and $Z$ portals, our framework also introduces the lepton portal and $Z'$ portal which originates from the broken $U(1)$ symmetry with a unique charge assignment. On the other hand, the $U(1)_D$ dark photon has a kinetic mixing term induced by the heavy exotic fermions. By providing a concrete UV origin for these interaction terms, our model offers additional information, such as the structure of the Yukawa couplings and the mass spectrum of the heavy mediators, allowing us to compare the relative strengths of each portal.

Below the dark color confinement scale, dark quarks form various dark hadrons, with dark pions naturally emerging as the lightest particles composed of dark quarks. The previously derived portal operators couple these dark pions to SM fermions, yielding signatures typical of fermionic axion-like particles (ALPs). Depending on the UV origin, these effective couplings can be either flavor-universal (if mediated via the gauge bosons) or flavor-dependent (if mediated via the scalar bosons). Our framework provides distinct benchmark scenarios for ongoing fermionic ALP searches.

Although we have adopted a strictly minimal scalar sector in this study, our setup can be easily extended to accommodate a wider range of interactions and associated phenomenological signatures. A broader exploration of the portal parameter space is therefore strongly encouraged. Moreover, while this work establishes a generalized framework, targeted analyses of specific dark color groups and particular portals remain crucial, demanding dedicated follow-up studies to fully unravel their phenomenological impact.

%~\\ - Goal ~\\

Ultimately, the present work is intended to offer a novel perspective on the structure of a strongly coupled dark sector and its portal with the SM. By providing a UV-complete framework, we hope to stimulate new directions in dark sector model building and related phenomenological studies. Beyond particle physics, this setup can also have implications for astrophysics and cosmology. For example, a strongly coupled dark sector naturally predicts sizable dark matter self-interactions \cite{Boddy:2014yra,Boddy:2014qxa,Cline:2013zca,Cline:2022leq,Garcia-Cely:2024ivo,Chung:2025wle}, offering a compelling resolution to the observed discrepancies in small-scale structure formation \cite{Spergel:1999mh,Dave:2000ar,Tulin:2017ara,Bullock:2017xww}. Furthermore, the rich dynamics of our model open up intriguing avenues for early-universe model building, such as darkogenesis \cite{Shelton:2010ta,Fujikura:2024jto,Girmohanta:2025wcq,Girmohanta:2026eqg,Girmohanta:2026ogq} and dark phase transitions \cite{Schwaller:2015tja,Nakai:2020oit,Fujikura:2023lkn}. We hope that this framework will ultimately shed light on some of the most profound dark matter puzzles, especially the dark matter-baryon coincidence problem \cite{Turner:1987pp}.

\acknowledgments

I thank Sudhakantha Girmohanta, Jong-Wan Lee, Huayang Song, Raymond Volkas, Giovani Dalla Valle Garcia, Giacomo Landini, Pedro Schwaller, and Adrian Carmona for many useful discussions. I would also like to express special thanks to the Mainz Institute for Theoretical Physics (MITP) of the Cluster of Excellence PRISMA+ (Project ID 390831469), for its hospitality and support during the completion of this work. I am supported by IBS under the project code, IBS-R018-D1.

%\newpage

\appendix

\section{Fermion and Scalar charges under the $U(1)$ Symmetries}\label{App:U(1)}
In this Appendix, we show the $U(1)$ charges of the relevant fermions $\mathbf{A}$, $\overline{\mathbf{F}}$, and $\mathbf{S'}$, as well as the scalar $\mathbf{\Phi}$, in matrix form. Starting from the initial $U(1)$ symmetries:

\paragraph{(1) $U(1)_N\subset SU(N+5)$ defined in eq. \eqref{GammaVEV}}
\begin{align}\label{Q_N}
Q_N(\mathbf{A}) =
\begin{pmatrix}
10      &   5-N   &  5-N \\
   &   -2N     &  -2N      \\
   &       &  -2N    \\
\end{pmatrix}
~,\quad
Q_N(\overline{\mathbf{F}}) =-Q_N(\mathbf{\Phi}) =
\begin{pmatrix}
-5  &  -5\\ 
N  &  N   \\ 
N  &  N  \\
\end{pmatrix}
~,\quad
Q_N(\mathbf{S'}) =0
\end{align}

\paragraph{(2) $U(1)'_N\subset SU(N+1)'$ defined in eq. \eqref{U(1)'N}}
\begin{align}
Q'_N(\mathbf{A}) =0
~,\quad
Q'_N(\overline{\mathbf{F}}) = Q'_N(\mathbf{\Phi}) =
\begin{pmatrix}
-1  &  N  \\ 
-1  &  N  \\ 
-1  &  N  \\
\end{pmatrix}
~,\quad
Q'_N(\mathbf{S'}) =
\begin{pmatrix}
2  &   1-N    \\
   &   -2N       \\
\end{pmatrix}
\end{align}

\paragraph{(3) $U(1)''$}
\begin{align}
Q''(\mathbf{A}) =N+1
~,\quad
Q''(\overline{\mathbf{F}}) =-(N+3)
~,\quad
Q''(\mathbf{S'}) = N+5
~,\quad
Q''(\mathbf{\Phi}) = -2
\end{align}

In the intermediate state, we have an unbroken $U(1)$ symmetry given by:

\paragraph{(4) $U(1)''_N \subset U(1)'_N\times U(1)''$ defined in eq. \eqref{T''N}}
\begin{align}\label{Q''N}
&Q''_N(\mathbf{A}) =2N(N+1)
~,\quad
Q''_N(\overline{\mathbf{F}}) =(N+1)
\begin{pmatrix}
-(2N+5)  &  -N  \\ 
-(2N+5)  &  -N  \\ 
-(2N+5)  &  -N  \\
\end{pmatrix}\nonumber\\
&
Q''_N(\mathbf{S'}) =(N+1)(N+5)
\begin{pmatrix}
2  &   1    \\
   &   0       \\
\end{pmatrix}~,\quad
Q''_N(\mathbf{\Phi}) =(N+1)
\begin{pmatrix}
-5  &  N  \\ 
-5  &  N  \\ 
-5  &  N  \\
\end{pmatrix}
\end{align}

At low energies, two $U(1)$ symmetries remain unbroken. One is the SM hypercharge $U(1)_Y$ as an unbroken subgroup of $SU(N+5)$, and the other is the dark baryon number $U(1)_D$ as an unbroken combination of $U(1)_N\times U(1)'_N\times U(1)''$:

\paragraph{(5) $U(1)_Y$ Hypercharge $\subset SU(N+5)$}
\begin{align}
Y(\mathbf{A}) =
\begin{pmatrix}
0      &   -\frac{1}{3}   &  \frac{1}{2} \\
   &   -\frac{2}{3}      &  \frac{1}{6}      \\
   &       &  1    \\
\end{pmatrix}
~,\quad
Y(\overline{\mathbf{F}}) = -Y(\mathbf{\Phi}) =
\begin{pmatrix}
0  &  0\\ 
\frac{1}{3}  &  \frac{1}{3}   \\ 
-\frac{1}{2}  &  -\frac{1}{2}  \\
\end{pmatrix}
~,\quad
Y(\mathbf{S'}) =0
\end{align}

\paragraph{(6) $U(1)_D$ Dark Baryon Number defined in eq. \eqref{U(1)D}}
\begin{align}
D(\mathbf{A}) =\frac{1}{N}
\begin{pmatrix}
2      &   1   &  1 \\
   &   0      &  0      \\
   &       &  0   \\
\end{pmatrix}
\,,~
D(\overline{\mathbf{F}}) =\frac{1}{N}
\begin{pmatrix}
-2  &  -1\\ 
-1  &  0   \\ 
-1  &  0  \\
\end{pmatrix}
\,,~
D(\mathbf{S'}) =\frac{1}{N}
\begin{pmatrix}
2      &   1    \\
   &   0       \\
\end{pmatrix}
\,,~
D(\mathbf{\Phi}) =\frac{1}{N}
\begin{pmatrix}
0  &  1\\ 
-1  &  0   \\ 
-1  &  0  \\
\end{pmatrix}
\end{align}

%\newpage

\section{Details of the $U(1)$ symmetry breaking chain}\label{App:U(1)SB}

$U(1)$ symmetries and their breaking play an important role in this model. In this Appendix, we detail the main $U(1)$ symmetry breaking chain
\begin{align}\label{U(1)SB}
U(1)_N\times U(1)'_N\times U(1)''~
\xrightarrow{~\Lambda_{\rm Seesaw}~} U(1)_N\times U(1)''_N~
\xrightarrow{~\Lambda_{\rm Diagonal}~} ~U(1)_D
\end{align} 
by first specifying the covariant derivative and gauge couplings as
\begin{align}
D=\partial-ig_NZ_N T_N-ig'_NZ'_N T'_N-ig''Z'' T''
\end{align}
where we do not write the $\mu$ index for convenience. Here, we follow the generators defined in section \ref{sec:ScalarSB} and the charges presented in Appendix \ref{App:U(1)}, and omit the normalization factors for convenience, as these do not affect our main results. The two symmetry breaking steps are discussed in detail below.~\\

\paragraph{(1) $U(1)'_N\times U(1)''\to U(1)''_N$ breaking by $\langle \mathbf{\Delta} \rangle=\frac{v_R}{\sqrt{2}}$ at the seesaw scale}~\\

The first breaking happens at the seesaw scale, giving a heavy $Z_R$ boson and a massless $Z''_N$ boson corresponding to the unbroken $U(1)''_N$ symmetry. The heavy $Z_R$ boson, with the subscript $R$ reflecting the fact that it is at the same scale as the right-handed neutrino mass $\mathbf{M}_R$, acquires a mass from the term
\begin{align}
\mathcal{L}=\frac{1}{2}v_R^2\left((-4N)g'_NZ'_N+(2N+10)g''Z''\right)^2=\frac{1}{2}M_{Z_R}^2Z_R^2~,
\end{align}
where the mass is
\begin{align}
M_{Z_R}=v_R\times \sqrt{(4Ng'_N)^2+((2N+10)g'')^2}~.
\end{align}
The massive $Z_R$ state and the massless $Z''_N$ state are given by
\begin{align}
Z_R={\rm cos}\varphi~ Z'_N-{\rm sin}\varphi~ Z''~,
\quad Z''_N={\rm sin}\varphi~ Z'_N+{\rm cos}\varphi~ Z''~,
\end{align}
where tan$\varphi={(N+5)g''}/{2Ng'_N}$.

The currents to which these gauge bosons couple can be extracted from
\begin{align}
iD=i\partial+
Z_R\left(g'_N{\rm cos}\varphi~T'_N-g''{\rm sin}\varphi~T''\right)+
Z''_N\left(g'_N{\rm sin}\varphi~T'_N+g''{\rm cos}\varphi~T''\right)~.
\end{align}
Since $Z_R$ has a relatively small phenomenological impact, we focus on the second term, which can be rewritten as
\begin{align}
iD(Z''_N)=
g''_NZ''_NT''_N~,\quad g''_N=\frac{g'_Ng''}{\sqrt{(2Ng'_N)^2+((N+5)g'')^2}}
\end{align}
where $T''_N=(N+5)\times T'_N+2N\times T''$ as derived in eq.~\eqref{T''N}. The corresponding charges of fermions and scalars are given in already eq.~\eqref{Q''N}.

%\newpage

\paragraph{(2) $U(1)_N\times U(1)''_N\to U(1)_D$ breaking by $\langle \mathbf{\Phi} \rangle=\frac{f}{\sqrt{2}}$ at the diagonal breaking scale}~\\

Next, we have $\mathbf{\Phi}$ triggering the $U(1)_N\times U(1)''_N$ breaking at a lower scale, which has greater phenomenological impact. The breaking introduces a heavy $Z'$ boson and a massless $A'$ boson corresponding to the unbroken $U(1)_D$ symmetry. Again, we start from the mass term induced by the VEV $\langle \mathbf{\Phi} \rangle$, which carries charges $5$ under $U(1)_N$ and $-5(N+1)$ under $U(1)''_N$, given by
\begin{align}
\mathcal{L}=\frac{1}{2}f^2\left(5g_NZ_N-5(N+1)g''_NZ''_N\right)^2=\frac{1}{2}M_{Z'}^2Z'^2~,
\end{align}
where the mass is
\begin{align}
M_{Z'}=f\times \sqrt{(5g_N)^2+((5N+5)g''_N)^2}~.
\end{align}
The massive $Z'$ state and the massless $A'$ state are given by
\begin{align}
Z'={\rm cos}\theta~ Z_N-{\rm sin}\theta~ Z''_N~,
\quad A'={\rm sin}\theta~ Z_N+{\rm cos}\theta~ Z''_N
\end{align}
where tan$\theta={(N+1)g''_N}/{g_N}$.

The current to which these gauge bosons couple can be extracted from
\begin{align}
iD=i\partial+
Z'\left(g_N{\rm cos}\theta~T_N-g''_N{\rm sin}\theta~T''_N\right)+
A'\left(g_N{\rm sin}\theta~T_N+g''_N{\rm cos}\theta~T''_N\right)
\end{align}
The second term gives the interaction between the dark photon $A'$ and the $U(1)_D$ current,
\begin{align}
iD(A')=
g_DA'T_D~,
\end{align}
where
\begin{align}\label{gD}
g_D=\frac{(N+1)(N+5)N~g_Ng''_N}{\sqrt{g_N^2+((N+1)g''_N)^2}}=
\frac{(N+1)(N+5)N~g_Ng'_Ng''}{\sqrt{(2Ng_Ng'_N)^2+((N+5)g_Ng'')^2+((N+1)g'_Ng'')^2}}
\end{align}
and
\begin{align}
T_D=\frac{(N+1)\times T_N + T''_N}{(N+1)(N+5)N}=
\frac{(N+1)\times T_N + (N+5)\times T'_N+2N\times T''}{(N+1)(N+5)N}~.
\end{align} 
The first term, on the other hand, describes the interaction between the $Z'$ and the particles carrying the corresponding charges, which is given by
\begin{align}
iD(Z')=Z'\left(\frac{1}{g_{Z'}}g_N^2 T_N-\frac{1}{g_{Z'}}(N+1){g''_N}^2 T''_N\right)=g_{Z'}Z'\left(T_N-N(N+5)\,{\rm sin}^2\theta~ T_D\right)~,
\end{align}
where we define $g_{Z'}\equiv\sqrt{g_N^2+((N+1)g''_N)^2}$.
Both the dark fermions and the SM fermions couple to the $Z'$. Since the SM fermions have $D=0$, their interactions are completely determined by the first term involving $T_N$, with the charge given in eq.~\eqref{Q_N}, i.e.,
\begin{align}
Q'(u^c)=Q'(q)=Q'(e^c)=-2N~,\quad 
Q'(d^c)=Q'(\ell)=N~,
\end{align}
which has some nontrivial phenomenological implications.

\bibliographystyle{jhepbst}
\bibliography{DM_Ref}{}

\end{document}